\documentclass[prl,aps,showpacs,floatfix,nofootinbib,twocolumn,letterpaper,reprint]{revtex4-1}
\usepackage{epsfig}
\usepackage{amssymb}
\usepackage{amsmath}
\usepackage{amsfonts}
\usepackage{color,graphicx}
\usepackage{bm}

\definecolor{violet}{RGB}{100,0,200}

\newcommand{\ssec}[1]{\emph{#1}.---}
\DeclareMathOperator{\Tr}{Tr}

\begin{document}
\title{Pseudogap effects in the strongly correlated regime of the two-dimensional Fermi gas}
\date{\today}
\author{S. Ramachandran$^1$}
\author{S. Jensen$^2$}
\author{Y. Alhassid$^1$}
\affiliation{$^1$Center for Theoretical Physics, Sloane Physics Laboratory, Yale University, New Haven, Connecticut 06520, USA
\\
$^2$Department of Physics, University of Illinois at Urbana-Champaign, Urbana, Illinois 61801, USA}
\begin{abstract}
The two-species Fermi gas with attractive short-range interactions in two spatial dimensions provides a paradigmatic system for the understanding of strongly correlated Fermi superfluids in two dimensions. It is known to exhibit a BEC-BCS  crossover as a function of  $\ln(k_F a)$, where $a$ is the scattering length, and to undergo a Berezinskii-Kosterlitz-Thouless superfluid transition below a critical temperature $T_c$. However, the extent of a pseudogap regime in the strongly correlated regime of $\ln(k_F a)\sim 1$, in which pairing correlations persist above $T_c$, remains largely unexplored with controlled theoretical methods. Here we use finite-temperature auxiliary-field quantum Monte Carlo (AFMC) methods on discrete lattices in the canonical ensemble formalism to calculate thermodynamical observables in the strongly correlated regime. We extrapolate to continuous time and the continuum limit to eliminate systematic errors and present results for particle numbers ranging from $N=42$ to $N=162$. We estimate $T_c$ by a finite-size scaling analysis, and observe clear pseudogap signatures  above $T_c$ and below a temperature $T^*$ in both the  spin susceptibility and free-energy gap. We also present results for the contact, a fundamental thermodynamic property of quantum many-body systems with short-range interactions. 
 \end{abstract}

\maketitle

\ssec{Introduction}  Cold atomic Fermi gases are of great interest in diverse areas of physics in part because they provide a well-defined paradigm of strongly correlated Fermi superfluids. They have been the subject of intensive experimental and theoretical studies. Of particular interest is the two-species uniform Fermi gas with attractive short-range interactions, whose strength can be controlled experimentally through a Feshbach resonance.  

The interaction strength is characterized by the two-particle $s$-wave scattering length $a$. In three spatial dimensions (3D), this system makes a crossover at low temperatures from a Bose-Einstein condensate (BEC) regime of weakly interacting dimers for $(k_F a)^{-1} \to  \infty$ to a Bardeen-Cooper-Schrieffer (BCS) regime at  $(k_F a)^{-1} \to -\infty$ ($k_F$ is the Fermi wavenumber). 
The physics of the interacting Fermi gas in two spatial dimensions (2D) differs qualitatively from that in 3D. Unlike the 3D case, where a two-particle bound state is formed when the interaction is sufficiently attractive, there is a bound state for arbitrarily weak attractive interactions~\cite{Randeria1989, Brodsky2006}. There is still a BEC to BCS crossover~\cite{Bertaina2011}  as a function of the scattering parameter $\eta = \ln(k_F a)$~\cite{Levinsen2015}, with the BEC and BCS limits corresponding to $\eta \rightarrow-\infty$ and $\eta\rightarrow \infty$, respectively.

While both the 2D and 3D systems undergo a superfluid transition below a critical temperature $T_c$, in 2D this phase transition does not have a non-vanishing condensate fraction with off-diagonal long-range order as in the 3D case, but instead exhibits a quasi-long-range order with algebraic decay of correlations in the superfluid regime. This 2D superfluid transition is known as a Berezinskii-Kosterlitz-Thouless (BKT) transition~\cite{Berezinskii1971, Berezinskii1972, Kosterlitz1973}. 

The 2D BEC-BCS crossover and BKT transition have been intensively studied both experimentally~\cite{Frohlich2011, Vogt2012, Murthy2015, Ries2015, Boettcher2016, Fenech2016, Toniolo2017, Luciuk2017, Hueck2018, Murthy2018} and theoretically~\cite{Watanabe2013, Matsumoto2014, Bauer2014, Anderson2015, Marsiglio2015, Shi2015, Galea2016, Vitali2017,Madeira2017, Schonenberg2017, Mulkerin2018, Hu2019, Wu2020, Pascucci2020, Zhao2020, Zielinski2020, Mulkerin2020a, Mulkerin2020b, Wang2020, He2022}. The strongly correlated regime, $\eta=\ln(k_F a) \sim 1$, of the 2D BEC-BCS crossover presents a particular theoretical challenge. In this regime there is no controlled analytic approach, and controlled computational methods provide the most reliable results. At zero temperature, the diffusion Monte Carlo studies of Ref.~\cite{Bertaina2011} calculated the energy, the pairing gap, and the contact, and Refs.~\cite{Shi2015, Galea2016, Vitali2017,Zielinski2020} addressed pairing correlations. At finite temperature, lattice auxiliary-field Monte Carlo (AFMC) methods were used in Ref.~\cite{Anderson2015} to calculate the pressure, compressibility, and contact of the 2D crossover but without a continuum extrapolation.  The BKT critical temperature was recently calculated in Ref.~\cite{He2022} as a function of $\eta$ using AFMC methods on large lattices in the grand-canonical ensemble with a continuum extrapolation.  

An open problem is the extent of a pseudogap regime, in which signatures of pairing correlations survive above the critical temperature $T_c$ for supefluidity.  Measurements of the spectral function of a harmonically trapped gas~\cite{Feld2011} indicate a pairing gap at temperatures well above the critical temperature in the strongly correlated regime. More recent radio-frequency spectroscopy experiments in the normal phase revealed the presence of an energy gap in the spectrum, which, at $\eta \sim 1$, far exceeds the two-body binding energy~\cite{Murthy2018}. In Ref.~\cite{Bauer2014}, the pseudogap regime was studied by calculating the single-particle spectral function using the Luttinger-Ward self-consistent field theory approach. A pronounced depression in the single-particle density of states was found in the normal phase of the strongly correlated regime, whereas the same method applied in 3D to the unitary Fermi gas showed a substantially reduced pseudogap signature~\cite{Haussmann2009, Zwerger2016, Pini2019}.
While this self-consistent $T$-matrix method compares remarkably well to AFMC for the unitary Fermi gas~\cite{Wlaz?owski2013,Jensen2020a,Richie2020,Rammelmuller2021}, it is nevertheless an uncontrolled approximation and its accuracy in 2D has not been established. 

In this work, we calculate thermodynamic properties of the 2D Fermi gas across the superfluid transition in its strongly correlated regime and, in particular, explore pseudogap effects above $T_c$. We employ lattice auxiliary-field quantum Monte Carlo (AFMC) methods in the canonical ensemble~\cite{Alhassid2017, Jensen2019, Jensen2020a} and extrapolate to continuous time and the continuum limit, thus eliminating any systematic errors and yielding results that are accurate up to statistical errors for a given particle number $N$.  This approach has the advantage of being a well-controlled computational method.

We estimate the critical temperature of the BKT transition using a finite-size scaling analysis of the largest eigenvalue of the two-body density matrix. We identify two pseudogap signatures above the critical temperature: (i) the suppression of the spin susceptibility above $T_c$ and below a temperature $T^*$ (also called the spin gap), and (ii) the increase in a model-independent free-energy gap with decreasing temperature within the spin gap regime. The calculation of the free-energy gap requires the use of canonical-ensemble AFMC and particle-reprojection~\cite{Alhassid1999, Jensen2023} methods. We find that the pseudogap regime is broad at a coupling of $\eta = 1$ and its width decreases with the coupling $\eta$ as the system approaches its BCS regime. 
We also calculate Tan's contact, a fundamental quantity for quantum many-body systems with short-range interactions that describes the short-distance pair correlations of opposite spin particles~\cite{Tan2008,Werner2012}. We find that the contact increases as the temperature decreases within the pseudogap regime.

\ssec{Finite-temperature canonical ensemble AFMC}  
We briefly describe the auxiliary-field quantum Monte Carlo (AFMC) method we use in this work; for recent reviews see Refs.~\cite{Gubernatis2017, Alhassid2017}. We consider a system of $N$ spin $1/2$ fermions interacting with a contact interaction of strength $V_0$ on a finite area with periodic boundary conditions.  We discretize our system on a  lattice of size $N_L^2$ with $N_L$ points in each direction and a lattice spacing $\delta x$. The corresponding lattice Hamiltonian is then
\begin{equation}\label{latticehamiltonian}
\hat{H} = \sum_{\mathbf{k}, s_z} \epsilon_{\mathbf{k}} \hat{a}^\dagger _{\mathbf{k}, s_z} \hat{a} _{\mathbf{k}, s_z} + g \sum_{\mathbf{x}_i} \hat{n}_{\mathbf{x}_i , \uparrow} \hat{n}_{\mathbf{x}_i , \downarrow} \;,
\end{equation}
where $\hat{a}^{\dagger }_{\bf{k},\sigma}$ and $\hat{a}_{\bf{k},\sigma}$ are creation and annihilation operators for fermions with momentum ${\bf k}$ and spin projection $s_z$ and $\epsilon_{\mathbf{k}} = \frac{\hbar^2 \mathbf{k}^2}{2m}$ for a quadratic dispersion. Here $g=V_0 / (\delta x)^2$ is the coupling strength, chosen to reproduce the physical scattering length on a lattice, and $\hat{n}_{\mathbf{x}_i,s_i}$ is the density operator of the fermions at position $\mathbf{x}_i$ and spin projection $s_z$.  The first sum on the r.h.s.~of Eq.~(\ref{latticehamiltonian}) is taken over the complete first Brillouin zone, and the second sum is taken over all lattice sites $\mathbf{x}_i$. 

We divide the inverse temperature $\beta$ into $N_\tau$ time slices of length $\Delta\beta = \beta/N_\tau$, and apply a symmetric Trotter-Suzuki decomposition to the imaginary-time propagator  $e^{-\beta \hat{H}} = \left( e^{-\frac{\Delta\beta}{2} \hat{H}_0} e^{-\Delta\beta \hat{V}} e^{-\frac{\Delta\beta}{2} \hat{H}_0} \right) ^{N_\tau} + \mathcal{O} [ \left ( \Delta \beta \right ) ^2 ]$, where $\hat{H}_0$ is the kinetic energy and $\hat V$ is the interaction term in the Hamiltonian (\ref{latticehamiltonian}). Rewriting $\hat{V} = g \sum_{\mathbf{x}_i}(\hat{n}_{\mathbf{x}_i} ^2 - \hat{n}_{\mathbf{x}_i})/2$, we use a Hubbard-Stratonovich transformation to decouple the two-body interaction introducing auxiliary fields $\{ \mathbf{\sigma}_{\mathbf{x}_i} (\tau_n)\}$ at each lattice site $\mathbf{x}_i$ and time slice $\tau_n$.  For $g<0$, the thermal propagator can then be written in the form
\begin{equation}\label{HSThermal}
e^{-\beta \hat{H}} = \int D[\mathbf{\sigma}] G_\mathbf{\sigma} \hat{U}_\sigma + \mathcal{O} \left [(\Delta\beta)^2 \right ]
\end{equation}
where $D[\mathbf{\sigma}] = \prod_{n=1} ^{N_\tau}  \prod_{\mathbf{x}_i} \left ( \sqrt{\Delta\beta |g|/2\pi} \right ) \textup{d}\sigma_{\mathbf{x}_i} (\tau_n)$ is the integration measure, $G_\mathbf{\sigma}=e^{-\Delta\beta |g| \sum_{i, n} \sigma^2 _{\mathbf{x}_i} (\tau_n) / 2}$ is a Gaussian weight, and $\hat{U}_\sigma=  \prod_n e^{-\Delta\beta \hat{H}_0 / 2} e^{-\Delta\beta \hat{h}_\sigma (\tau_n)} e^{-\Delta\beta \hat{H}_0 / 2}$ is the time-ordered one-body thermal propagator for a given auxiliary-field configuration, where $\hat{h}_\sigma(\tau_n)= g\sum_{\mathbf{x}_i} \mathbf{\sigma}_{\mathbf{x}_i} (\tau_n) \hat{n}_{\mathbf{x}_i} - g\hat{N}/2$ is the corresponding one-body Hamiltonian at any given time slice $\tau_n$. Each integral over a continuous auxiliary field $ \mathbf{\sigma}_{\mathbf{x}_i} (\tau_n)$ is discretized by a three-point Gaussian quadrature~\cite{Dean1993}.  

In canonical-ensemble AFMC we are interested in the calculation of thermal observables at fixed particle numbers $N_\uparrow,N_\downarrow$
\begin{equation}\label{observables}
\!\!\!\langle \hat{O} \rangle_{N_\uparrow,N_\downarrow} = \frac{\Tr_{N_\uparrow,N_\downarrow}(\hat{O} e^{-\beta\hat{H}})}{\Tr_{N_\uparrow,N_\downarrow} (e^{-\beta\hat{H}})} = \frac{\int D[\mathbf{\sigma}] \langle \hat{O} \rangle_\sigma G_\mathbf{\sigma} \Tr_{N_\uparrow,N_\downarrow}\hat{U}_\sigma}{\int D[\mathbf{\sigma}] G_\mathbf{\sigma} \Tr_{N_\uparrow,N_\downarrow}\hat{U}_\sigma}
\end{equation}
where $\langle \hat{O} \rangle_\sigma = \frac{\Tr_{N_\uparrow,N_\downarrow}(\hat{O} \hat{U}_\sigma)}{\Tr_{N_\uparrow,N_\downarrow}\hat{U}_\sigma}$ is the expectation value of $\hat{O}$ for a given field configuration. The traces at fixed particle numbers are calculated by using projection operators $\hat{P}_{N_{s_z}}$, i.e., $\Tr_{N_\uparrow,N_\downarrow} X= \Tr(\hat P_{N_\uparrow}  \hat P_{N_\downarrow}\hat X)$.  
For a finite number $N_s$ of single-particle states (for given spin projection $s_z$), the particle-number projection can be represented exactly by a discrete finite Fourier sum~\cite{Ormand1994}
 \begin{equation}\label{particlenumberprojection}
\hat{P}_{N_{s_z}} = \frac{e^{-\beta \mu N_{s_z}}}{N_s} \sum_{m = 1} ^{N_s} e^{-i \varphi_m N_{s_z}} e^{\left ( \beta \mu + i \varphi_m \right ) \hat{N}_{s_z}} \;.
\end{equation}
where $\varphi_m = 2\pi m/N_s$ are quadrature points, and $\mu$ is a real chemical potential introduced to stabilize the Fourier sum. 
Using (\ref{particlenumberprojection}), the fixed particle-number traces on the r.h.s. side of Eq. (\ref{observables}) reduce to the calculation of the unrestricted traces involving the propagator $ e^{(\beta\mu + i\phi_m)\hat{N}}\hat{U}_\sigma$. Since the latter is a one-body propagator, these traces can be expressed in terms of the single-particle representation matrix $ \mathbf{U}_\sigma$ of the many-particle propagator $\hat U_\sigma$. For example
\begin{equation}\label{tracedeterminant}
\Tr \left [e^{(\beta\mu + i\varphi_m)\hat{N}}\hat{U}_\sigma \right] = \det \left [1 +e^{(\beta\mu + i\varphi_m)} \mathbf{U}_\sigma \right] \;.
\end{equation}

\begin{figure*}[t]
  \includegraphics[width=\textwidth]{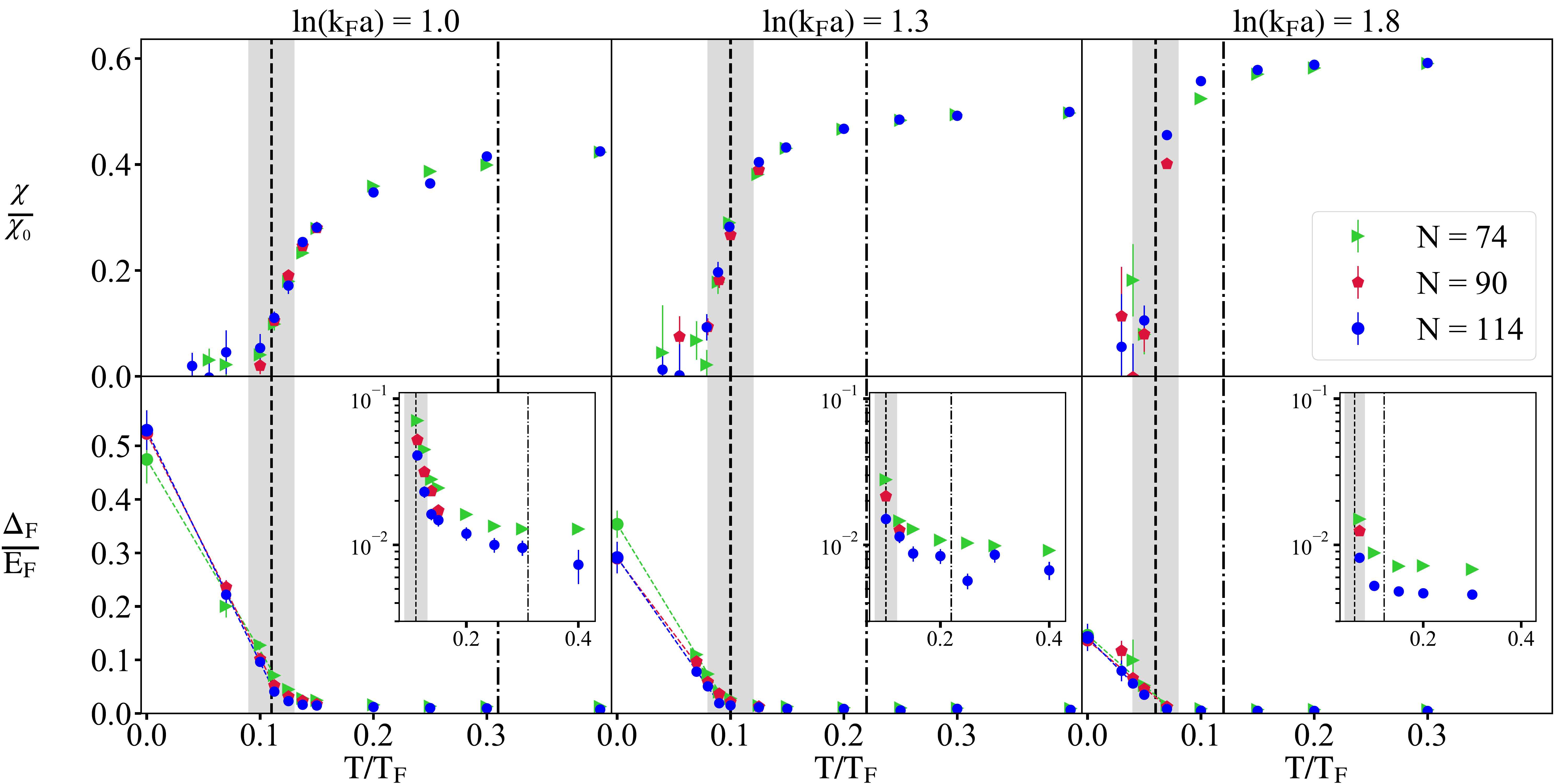}
 \caption{Spin susceptibility $\chi$ (top row) and free energy gap $\Delta_F$ (bottom row) vs.~temperature $T/T_F$ for couplings of $\eta = 1.0$ (left column),  $\eta = 1.3$ (middle column), and $\eta = 1.8$ (right column) for particle numbers $N=74, 90$ and $114$. The black dashed line with the gray band is $T_c$, and the dotted-dashed line is $T^*$.  $\chi$ is measured in units of the free gas zero-temperature susceptibility $\chi_0$ and $\Delta_F$ is measured in units of the free Fermi gas energy $E_F$.
  The insets in the bottom row show the free energy gap on a logarithmic scale (see text).}
  \label{fig:pairing}
\end{figure*}

In canonical-ensemble AFMC, we sample auxiliary-field configurations according to the positive-definite distribution $G_\sigma | \Tr_{N_\uparrow,N_\downarrow} \hat U_\sigma|$ and use them to estimate the thermal canonical expectation values of observables following Eq.~(\ref{observables}). 

We take the continuous time limit $\Delta\beta\rightarrow 0$ by performing calculations for several values of $\Delta\beta$, and extrapolation in $(\Delta\beta)^2$. We then take the continuum limit by performing calculations for a range of lattices containing as many as $29^2$ points, and performing a linear extrapolation in the filling factor $\nu = N / N_{L}^2 \rightarrow 0$~(see the Supplemental Material for detailed fits). The continuum limit requires large lattice calculations, which were made possible through the use of two algorithms: (i) a stable diagonalization method that reduces the $N_s^4$ scaling of the stabilization of canonical-ensemble AFMC to $N_s^3$~\cite{Gilbreth2015}, and (ii) a controlled truncation of the single-particle model space that further reduces the $N_s^3$ scaling to essentially $N_s N^2$ for most parts of the algorithm~\cite{Gilbreth2021}, offering a dramatic improvement for $N \ll N_s$.  A similar method was implemented in Ref.~\cite{He2019}.  

\ssec{Results} We present canonical-ensemble AFMC results for three different couplings in the strongly correlated regime of the 2D BEC-BCS crossover: $\eta=\ln(k_F a) = 1.0, 1.3$, and $1.8$. For each of these couplings, we carried out calculations for $N =42, 74, 90, 114$, and $162$ particles.

 We estimate the critical temperature $T_c$ of the transition to superfluidity for a given $\eta$ using the phenomenological finite-size scaling approach of Refs.~\cite{Nightingale1982, Santos1981} which is suitable for the BKT universality class. In this approach we scale the largest eigenvalue $\lambda_{\rm max}$ of the two-body density matrix $\langle a^\dagger _{\mathbf{k}_1 , \uparrow} a^\dagger _{\mathbf{k}_2 , \downarrow} a _{\mathbf{k}_3 , \downarrow} a _{\mathbf{k}_4 , \uparrow} \rangle$. For details, see the Supplemental Material~\cite{supp}. 

 We find $T_c = 0.11(2)\, T_F$ at $\eta = 1.0$, $T_c = 0.09(2)\, T_F$ at $\eta = 1.3$, and $T_c = 0.06(2)\,T_F$ at $\eta = 1.8$.  These estimates agree within error bars with the results of Ref.~\cite{He2022} obtained by a different method. They are lower than the experimental estimates of $T_c$ in Ref.~\cite{Ries2015}.

\noindent (i) Spin susceptibility: We calculate the spin susceptibility $\chi$ using
\begin{equation}
\chi = \frac{\beta}{V} \langle (N_\uparrow - N_\downarrow) ^2 \rangle \;.
\end{equation}
In the calculation of $\chi$, we project only on the total number of particles $N = N_\uparrow + N_\downarrow$, as opposed to the two-species projection used for other observables. 

The spin susceptibility is suppressed by pairing correlations. Our AFMC results for $\chi$ in units of $\chi_0$, the zero-temperature susceptibility of the non-interacting Fermi gas, are shown in the top row of Fig.~\ref{fig:pairing} as a function of $T/T_F$.  We observe the suppression of $\chi$ as the temperature is decreased below $T_c$. However, we also see moderate suppression of $\chi$ above $T_c$ but below a temperature scale $T^*$, known as a spin gap. We define $T^*$ to be the temperature at which $\chi/\chi_0$ reaches 95\% of  its maximal value. This spin gap regime becomes narrower with increasing $\eta$. We estimate $T^* = 0.31\,T_F, 0.22\,T_F$ and $0.12\,T_F$ for $\eta = 1, 1.3$ and $1.8$, respectively.

\noindent (ii) Free energy gap:
We define the free energy gap by
\begin{multline}
\Delta_F = [2F\left(N/2-1,N/2\right)-F\left (N/2-1,N/2-1\right)\\-F\left(N/2,N/2\right)]/2 \;,
\end{multline}
where $F(N_\uparrow, N_\downarrow)$ is the free energy for the system at $N_\uparrow$ spin-up particles and $N_\downarrow$ spin-down particles.  To calculate $\Delta_F$, we rewrite it as
\begin{equation}
\bold{\bigtriangleup}_{F} = -k_B T \ln \left(\frac{Z_{N/2,N/2+1}}{Z_{N/2,N/2}} \frac{Z_{N/2,N/2+1}}{Z_{N/2+1,N/2+1}}\right)  \;,
\end{equation}
where $Z_{N_{\uparrow},N_{\downarrow}}$ is the partition function for $N_{\uparrow}$ spin-up particles and $N_{\downarrow}$ spin-down particles.
Each partition function ratio can then be calculated using particle-number reprojection~\cite{Alhassid1999, Jensen2023}, e.g.,
\begin{equation}
\frac{Z_{N/2,N/2+1}}{Z_{N/2,N/2}}= \left\langle \frac{ {\rm Tr}_{N/2,N/2+1} U_\sigma} {{\rm Tr}_{N/2,N/2}  U_\sigma} \right\rangle_W \;,
\end{equation}
where we have introduced the notation $\langle X_\sigma \rangle_W \equiv {\int D[\sigma] W_\sigma X_\sigma /
\int  D[\sigma] W_\sigma}$ with $W_\sigma= G_{\sigma}  {\rm Tr}_{N/2,N/2}  U_\sigma$ a positive definite weight function for $N_{\uparrow} =N_{\downarrow} = N/2$.

\begin{figure}[bth]
    \centering
    \includegraphics[width=0.45\textwidth]{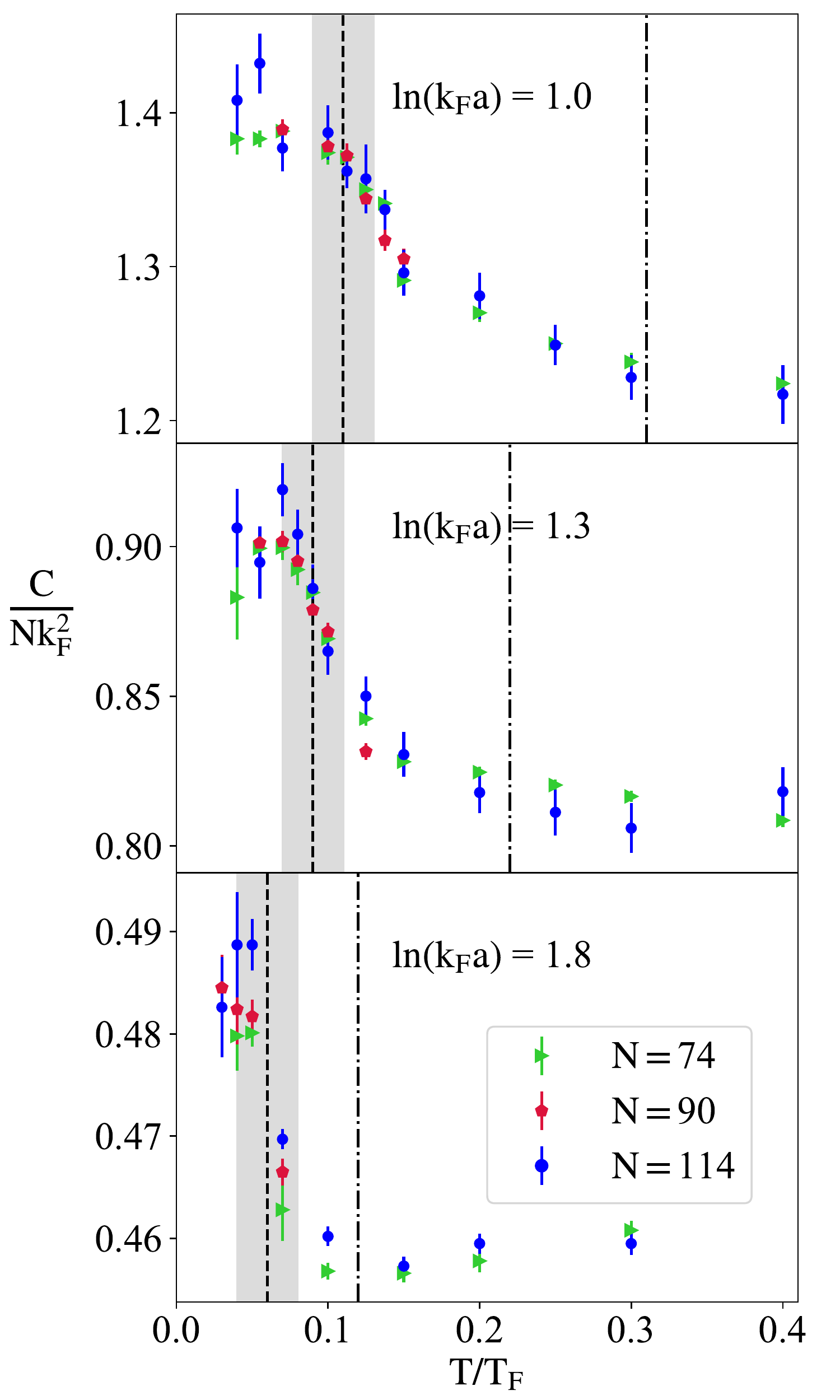}
    \caption{The contact $C$ (in units of $Nk_F^2$) vs.~temperature for couplings of $\eta = \ln(k_F a)=1.0, 1.3$, and $1.8$ for particle numbers $N=74, 90$ and $114$. The dashed black line with the gray band represents $T_c$, and the dotted-dashed line is $T^*$. }
    \label{fig:contact}
\end{figure}

We show our results for $\Delta_F$ as a function of $T/T_F$ in the bottom row of Fig.~\ref{fig:pairing}. Using a linear extrapolation at low temperatures, we determine the zero-temperature energy staggering pairing gap~\cite{Carlson2003,Gezerlis2008} $\Delta$ from the finite-temperature $\Delta_F$.  For additional details, see the Supplemental Material~\cite{supp}.  A direct AFMC calculation of $\Delta$ is challenging due to a sign problem introduced by the spin imbalance.  

We observe that the free energy gap $\Delta_F$  is suppressed above $T_c$.
To see more closely the behavior of $\Delta_F$ above $T_c$, we show $\Delta_F$ in the insets using a logarithmic scale.
In the spin gap regime between $T_c$ and $T^*$ we observe an increase in $\Delta_F$ with decreasing temperature due to pairing correlations. We identify this behavior of $\Delta_F$ as a pseudogap signature that correlates with the suppression of the spin susceptibility $\chi$. 

\noindent (iii) Contact:
The contact $C$ describes the short-range correlations between particles of opposite spin and is defined by
\begin{equation}
\int d^2 R \ g^{(2)} _{\uparrow\downarrow} \left ( \mathbf{R} + \frac{\mathbf{r}}{2} , \mathbf{R} - \frac{\mathbf{r}}{2} \right ) \underset{r\to 0}{\sim} \frac{C}{(2\pi)^2} \ln ^2 r \;,
\end{equation}
where $g^{(2)} _{\uparrow\downarrow} \left ( \mathbf{r}_\uparrow,\mathbf{r}_\downarrow \right ) = \langle \hat{n}_\uparrow \left ( \mathbf{r}_\uparrow \right ) \hat{n}_\downarrow \left ( \mathbf{r}_\downarrow \right ) \rangle$ is the two-body correlation function with $\hat{n}_\uparrow \left ( \mathbf{r}_\uparrow \right)$ and $\hat{n}_\downarrow \left ( \mathbf{r}_\downarrow \right)$ the density of spin up and spin down particles, respectively. The contact is of key interest due to Tan's relations~\cite{Tan2008,Werner2012}, which relate the contact to several different properties of the interacting Fermi gas.  In particular, the contact characterizes the tail of the momentum distribution  $n_{\sigma} (\mathbf{k}) \sim C/k^4$
in the limit $k\rightarrow\infty$. It can also be obtained from the derivative of the thermal energy with respect to the 
coupling parameter
\begin{equation}
\frac{\hbar^2 C}{2\pi m} = \frac{d E}{d \ln a} \;.
\end{equation}
 In the lattice simulations we calculate $C$ from the thermal expectation value $\langle V \rangle$ of the potential energy~\cite{Jensen2020b} 
 \begin{equation}
C = \frac{m^2}{\hbar^4} V_0 \langle \hat{V} \rangle \;.
\end{equation}
 We show our results for the contact (in units of $N k_F^2$) as a function of temperature for the three coupling values in Fig.~\ref{fig:contact}.  The contact increases rapidly with lowering $T$ in the vicinity of $T_c$. This increase occurs within a narrower temperature range as the coupling parameter $\eta$ increases. In general, the contact decreases with increasing $\eta$. In the pseudogap regime between $T_c$ and $T^*$, we observe a monotonic increase in the contact with decreasing $T$. 
 
\ssec{Conclusion and outlook}  We explored thermal properties of the two-species spin-balanced Fermi gas with short-range interactions in two spatial dimensions in the strongly correlated regime $\eta=\ln(k_Fa) \sim 1$. We used canonical-ensemble AFMC methods on discrete lattices and extrapolated our results to continuous time and the continuum limit. We estimated the critical temperature $T_c$ for the superfluid transition and calculated several thermodynamic observables across the superfluid phase transition for several values of $\eta$. We observed pseudogap signatures in a temperature regime $T_c < T < T^*$, including a suppression of the spin susceptibility and an increase in the free energy gap with decreasing temperature.  Our AFMC results for the spin susceptibility and free-energy gap are the first controlled calculation of observables that probe the pseudogap regime in two spatial dimensions and thus provide an accurate benchmark for experiment.
We also calculated Tan's contact and found it to increase monotonically with decreasing temperature in the pseudogap regime. 
In future AFMC studies, it will be interesting to calculate the spectral function of the 2D strongly interacting Fermi gas across the BEC-BCS crossover as a dynamical probe of the pseudogap regime.  
 
\ssec{Acknowledgements}  This work was supported in part by the U.S. DOE grants No.~DE-SC0019521 and No.~DE-SC0020177. The calculations used resources of the National Energy Research Scientific Computing Center (NERSC), a U.S. Department of Energy Office of Science User Facility operated under Contract No.~DE-AC02-05CH11231.  We thank the Yale Center for Research Computing for guidance and use of the research computing infrastructure.


%

 
\setcounter{figure}{0} 

\onecolumngrid
 \newpage
\begin{center}
{\large \bf Supplemental Material: Pseudogap effects in the strongly interacting regime of the two-dimensional Fermi gas}
\end{center}
\vspace*{7 mm} 
\twocolumngrid

We discuss a number of technical details including the continuum extrapolations and finite-size scaling.  We present additional results for the zero-temperature pairing gap, the thermal energy,  and the single-particle momentum distribution. 

\subsection{Extrapolations}

The lattice AFMC calculations are carried out by dividing $\beta$ into $N_t$ time slices of finite length $\Delta \beta$ and for discrete lattices, and it is necessary to take the  continuous time limit  $\Delta\beta \to 0$ and the continuum limit $\nu \to 0$ (where $\nu$ is the filling factor). In the following we  demonstrate how these limits are calculated.

\subsubsection{Continuous time limit}

\begin{figure*}[h!]
    \includegraphics[width= \textwidth]{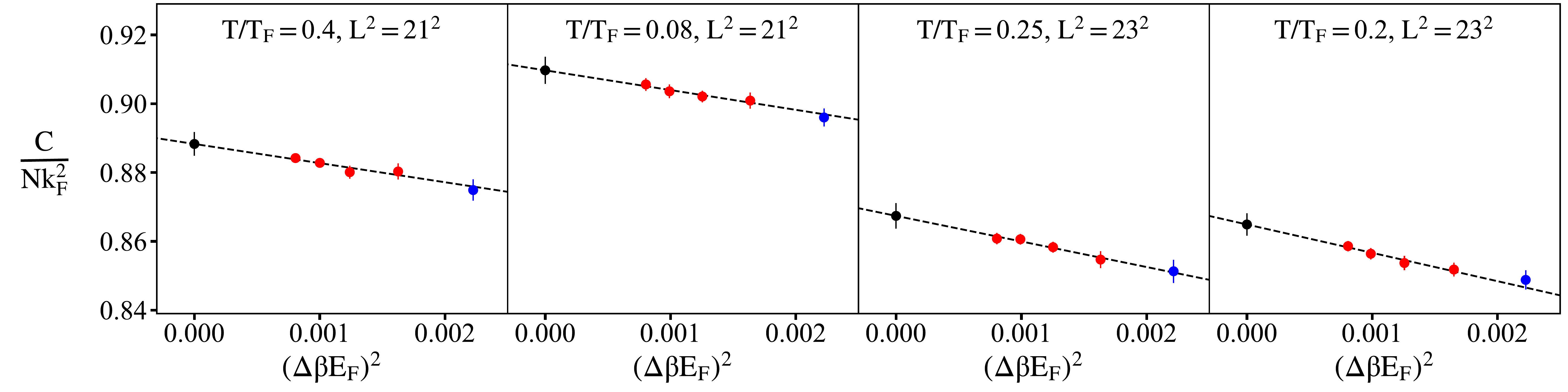}
    \caption{The contact $C$ (in units of $N k_F^2$) versus  the dimensionless parameter $(E_F \Delta\beta)^2$  for a coupling parameter of  $\ln(k_F a) = 1.3$ and $N=114$ particles. The dashed lines are linear extrapolations in $(\Delta\beta)^2$ to find the limit $\Delta\beta \rightarrow 0$.  Only the points in red below a cutoff of $(\Delta\beta  E_F)^2  < 0.002$ are included in the fit, and the extrapolated results for $\Delta\beta = 0$ are shown in black solid circles. 
}
\label{fig:ContactDB}
\end{figure*}

There are two systematic errors in $\Delta\beta$.  One arises from the symmetric Trotter-Suzuki decomposition of $e^{-\beta \hat H}$ which is of order $(\Delta\beta)^2$, and a second arises from using a three-point Gaussian quadrature formula to evaluate the integral over each auxiliary field,  which is accurate to order $(\Delta\beta)^2$, 
We eliminate these systematic errors by performing a linear extrapolation in $(\Delta\beta)^2$ to find the limit $\Delta\beta\rightarrow 0$. We demonstrate these extrapolations for the contact in Fig.~\ref{fig:ContactDB}. We note that sufficiently small values of $\Delta\beta$ are necessary in order to reach the quadratic regime. 

For the  the maximal eigenvalue of the two-body density matrix and the thermal energy, we find only a weak dependence on $\Delta\beta$ for small values of $\Delta\beta$, in which case we perform a flat extrapolation (i.e., take an average value).

\begin{figure*}
\includegraphics[width=\textwidth]{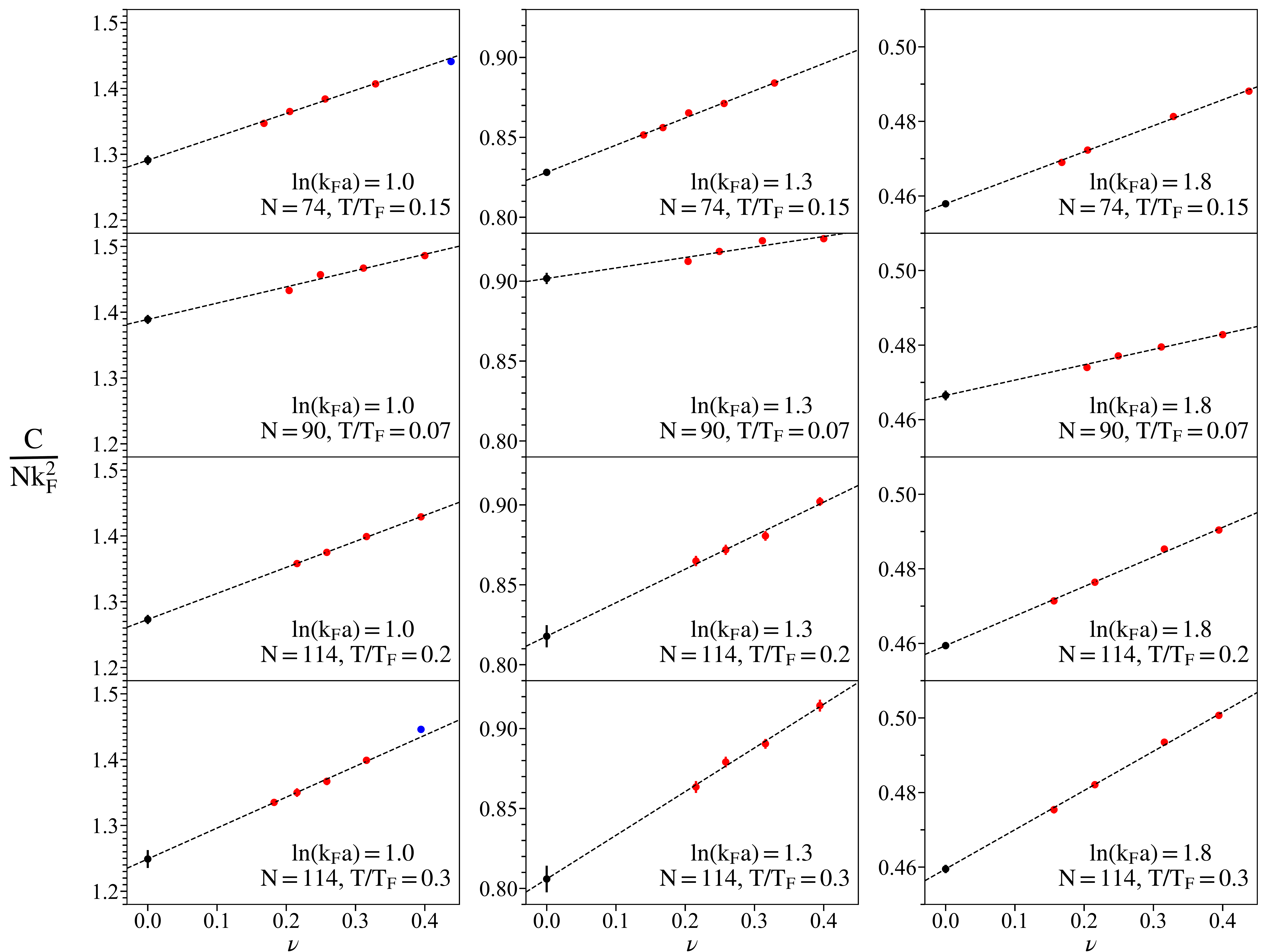}
    \caption{The $\Delta \beta=0$ contact $C$ (in units of $N k_F^2$)  versus the filling factor $\nu$  for $\ln(k_F a) = 1$ (left column), $\ln(k_F a) = 1.3$ (middle column), and  $\ln(k_F a) = 1.8$ (right column) for various temperatures and particle numbers.  The dashed lines are linear extrapolations in $\nu$ to determine the continuum limit at $\nu=0$. }
 \label{fig:ContactFilling}
\end{figure*}

\subsubsection{Continuum extrapolations}

\begin{figure}
    \includegraphics[width=0.5\textwidth]{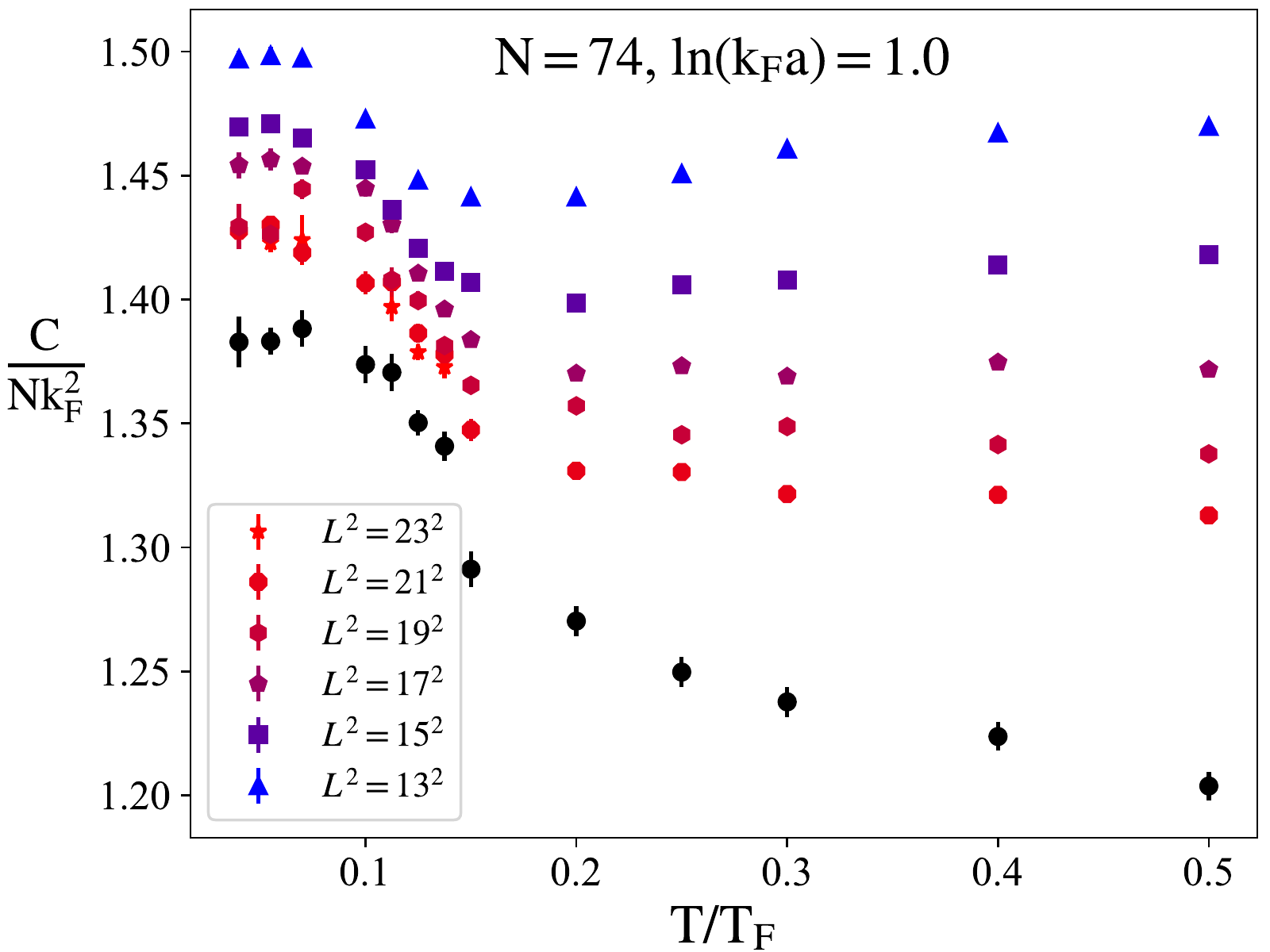}
    \caption{The contact $C$ versus $T/T_F$ for different finite lattice sizes ranging from $13^2$ ($\nu=0.44$) to $23^2$ ($\nu=0.14$).  The extrapolated continuum results at $\nu\rightarrow 0$ are shown by the black solid circles. We notice a strong dependence of the contact on the filling factor. }
    \label{fig:ContactExtrap}
\end{figure}

\begin{figure}[h!]
    \includegraphics[width=0.45\textwidth]{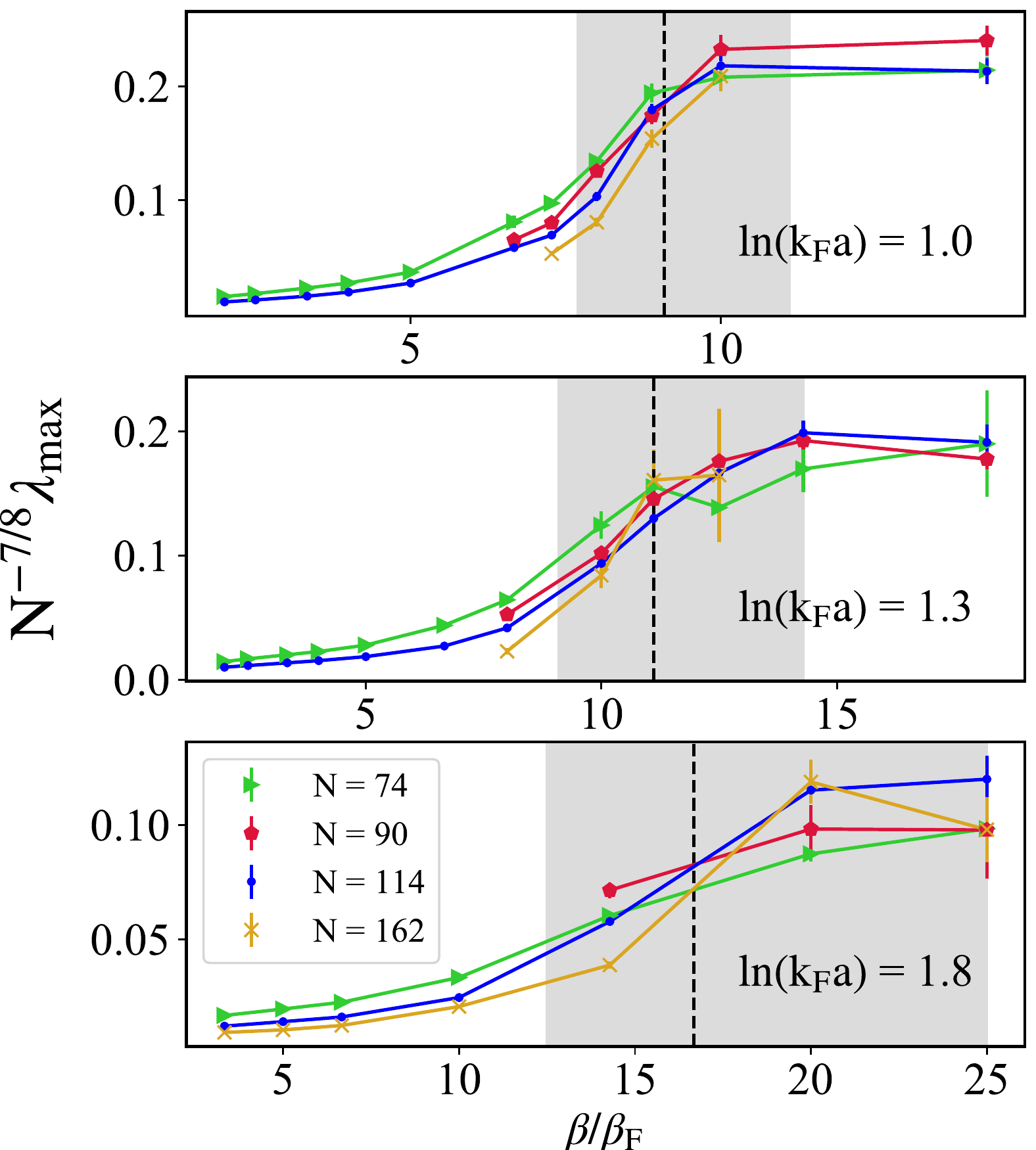}
    \caption{$\lambda_{\rm max}$ (see text) scaled by $N^{-7/8}$ as a function of $\beta/\beta_F$ at the three coupling values. We observe the merging  of curves with varying particle number in the critical region, marked by the gray bands with a black dashed line at the critical temperature $T_c$. The gray bands are used to estimate the errors in $T_c$.  
}
\label{fig:FSS}
\end{figure}

The continuum limit is obtained by taking the limit  $\nu=N/N_L^2 \rightarrow 0$, and is achieved by considering increasing lattice sizes. Thermal observables behave linearly in $\nu$ for small values of $\nu$ and we use linear extrapolations to extract the values of observables in the limit $\nu \rightarrow 0$. 

Examples of the linear extrapolations in $\nu$ are shown in Fig.~\ref{fig:ContactFilling} for the contact.  We find that this observable has a relatively strong dependence on $\nu$.  In Fig.~\ref{fig:ContactExtrap}, we show the contact (in units of $N k_F^2$) for $N=74$ and a coupling of $\eta=1$ as a function of $T/T_F$ for several lattice sizes $N_L^2$. The extrapolated  contact in the limit $\nu \rightarrow 0$ (black solid circles) show significant differences with the finite filling factor results even on a qualitative level. In particular,  the contact in the continuum limit above the critical temperature decreases with increasing temperature. In contrast, the spin susceptibility (not shown) exhibits only a weak dependence on $\nu$.

\subsection{Finite-size scaling}

We estimate the critical temperature $T_c$ of the transition to superfluidity through a finite-size scaling analysis  of the largest eigenvalue $\lambda_{\rm max}$ of the two-body density matrix $\langle a^\dagger _{\mathbf{k}_1 , \uparrow} a^\dagger _{\mathbf{k}_2 , \downarrow} a _{\mathbf{k}_3 , \downarrow} a _{\mathbf{k}_4 , \uparrow} \rangle$. We use the phenomenological renormalization group analysis of Refs.~\cite{Nightingale1982S, Santos1981S}. In the ordered phase $\lambda_{\rm max}$ scales as~\cite{Moreo1991S} 
\begin{equation} 
\lambda_{\rm max}  = L^{2 - \eta} f(L / \xi) \;,
\end{equation} 
where $L$ is the linear size of the system, $\xi$ is the correlation length,  and the exponent $\eta = 1/4$ at $T=T_c$~\cite{Kosterlitz1973S}. For a BKT transition, the correlation length $\xi\rightarrow\infty$ for $T\leq T_c$, 
and we expect the scaled curves $L^{-7/4} \lambda_{\rm max}$ across different values of $L$ to merge for $T\rightarrow T_c$~\cite{Paiva2004S} (rather than intersecting in the manner typical of the three-dimensional superfluid phase transition~\cite{Nightingale1982S}).
At constant density, the particle number $N \propto L^2$, and we scale $\lambda_{\rm max}$ according to $L^{-7/4}  \lambda_{\rm max}  \sim N^{-7/8} \lambda_{\rm max}$ for systems with different particle number $N$. 

We demonstrate this finite-size scaling analysis in Fig.~\ref{fig:FSS}, where  $N^{-7/8} \lambda_{\rm max}$ is shown as a function of $\beta/\beta_F$ for several values of $N$ at the three coupling values. We observe the merging of the different curves in the critical regime and use it to estimate $T_c$ and its associated error.  We find $T_c = 0.11(2)\, T_F$ at $\eta = 1.0$, $T_c = 0.09(2)\, T_F$ at $\eta = 1.3$, and $T_c = 0.06(2)\,T_F$ at $\eta = 1.8$.

\subsection*{Zero-temperature pairing gap}

\begin{figure}
    \centering
    \includegraphics[width=0.5\textwidth]{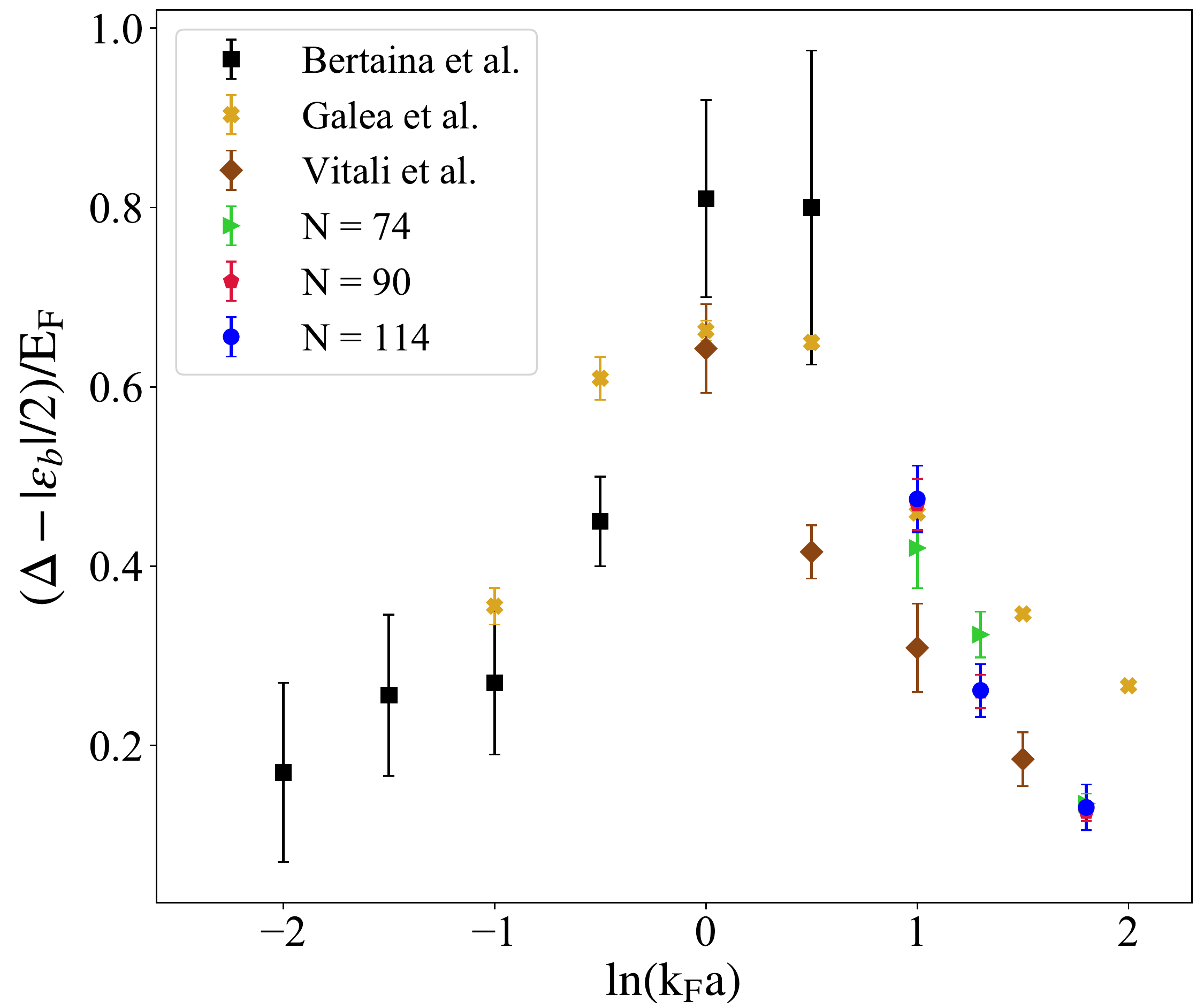}
    \caption{$\Delta -|\epsilon_b|/2$ (see text) as a function of the coupling parameter $\eta=\ln(k_F a)$.  Our results, based on extrapolations of the AFMC free energy gaps to $T=0$, are shown by the green, red and blue symbols for $N=74, 90$ and $114$, respectively. The  solid squares, x's and diamonds are, respectively, the diffusion Monte Carlo results of Refs.~\cite{Bertaina2011S,Galea2016S,Vitali2017S}.  }
    \label{fig:TZeroGap}
\end{figure}

At $T=0$ the free energy gap coincides with the energy staggering pairing gap.  The statistical errors on the free-energy gap are relatively small and at low temperatures we can use a linear extrapolation to determine the energy gap $\Delta$ at $T=0$.  In Fig.~\ref{fig:TZeroGap} we compare our results for $\Delta - |\epsilon_b/2|$ (here $\epsilon_b = -4\hbar^2 / (m a^2 e^{2\gamma})$ is the two-particle binding energy with  $\gamma\approx0.577$ being the Euler-Mascheroni constant) for several values of the particle number $N$ with the diffusion Monte Carlo results  of Ref.~\cite{Bertaina2011S} (solid squares),  Ref.~\cite{Galea2016S}  (x's) and Ref.~\cite{Vitali2017S} (solid diamonds). 

\subsection*{Thermal energy}

\begin{figure*}[t]
    \centering
    \includegraphics[width=0.93 \textwidth]{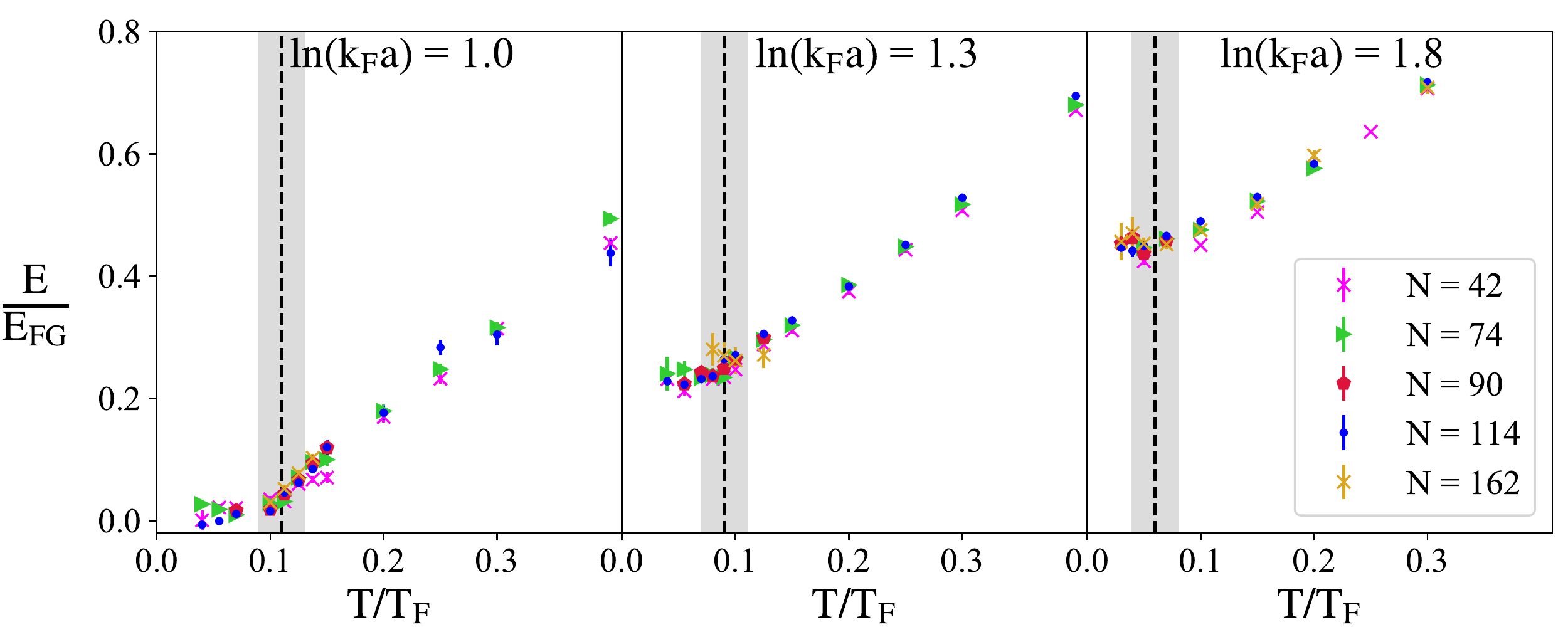}
    \caption{ Thermal energy versus $T/T_F$  for particle numbers ranging from 42 to 162 and for coupling parameters of $\ln (k_F a) = 1.0$, $\ln (k_F a) = 1.3$, and $\ln (k_F a) = 1.8$ We notice that the energy below $T_c$ increases with increasing $\ln (k_F a)$, in agreement with Ref.~\cite{Bertaina2011S}.}
    \label{fig:Energy}
\end{figure*}

\begin{figure*}[t!]
    \includegraphics[width=0.93 \textwidth]{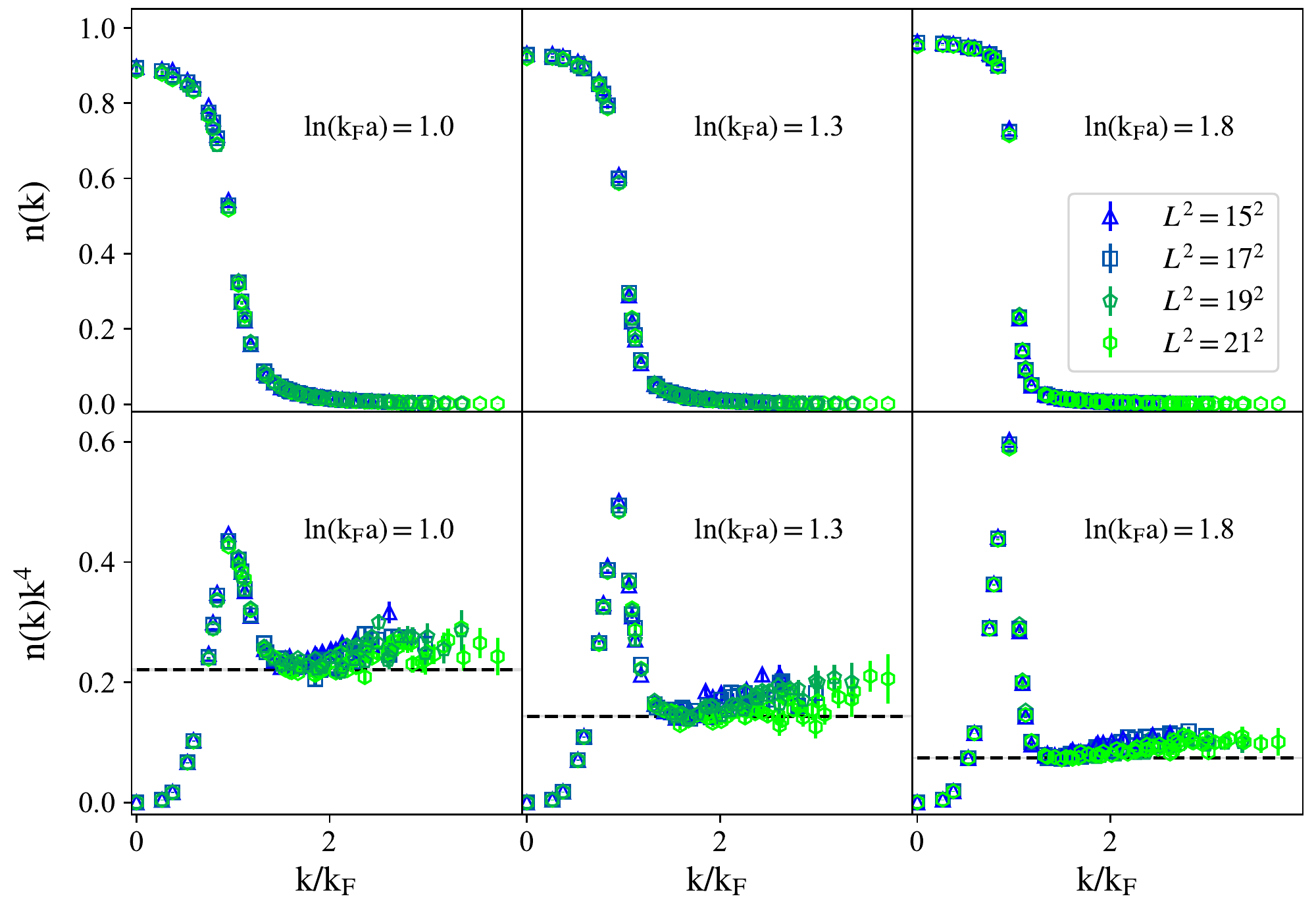}
    \caption{Left column: single-particle momentum distribution $n (k)$ for various couplings and lattice sizes at $T = 0.07\ T_\textup{F}$ and $N=90$.
    Right column: as in the left column but showing $n (k) k^4$ versus $k$. The black dashed line is the continuum limit contact obtained from $\langle V \rangle$. }
    \label{fig:Nk}
\end{figure*}

In Fig.~\ref{fig:Energy} we show our results for the thermal energy, defined as the expectation value of the Hamiltonian, as a function of $T/T_F$ for $N=42, 74, 90, 114$ and $162$.  The thermal energy increases linearly with temperature above the critical temperature, and depends weakly on temperature below the critical temperature.
 Our results are consistent with the ground-state energies calculated in Ref.~\cite{Bertaina2011S}.

\subsection*{Single-particle momentum distribution}

Our AFMC results for the single-particle momentum distribution, $n(k)$, are shown in the top row of Fig.~\ref{fig:Nk} for $N=90$ and $T=0.07  \,T_F$. Of particular interest is the tail of the momentum distribution. According to Tan's universality relations~\cite{Tan2008S, Werner2012S}, $n (k) \rightarrow C/ k^{4}$ at large $k$, where $C$ is the contact. In the bottom row of Fig.~\ref{fig:Nk}, we compare $n (k) k^4$ with the value of $C$ extracted from the average potential energy (dashed lines) and find good agreement for the larger lattices. It is difficult to take the continuum limit extrapolation of the momentum distribution tail  because of edge effects. Thus, calculating the contact from $\langle V \rangle$ is a more effective method of obtaining the continuum results for the  contact.

\newpage 

\begin{thebibliography}{63}%
\makeatletter
\providecommand \@ifxundefined [1]{%
 \@ifx{#1\undefined}
}%
\providecommand \@ifnum [1]{%
 \ifnum #1\expandafter \@firstoftwo
 \else \expandafter \@secondoftwo
 \fi
}%
\providecommand \@ifx [1]{%
 \ifx #1\expandafter \@firstoftwo
 \else \expandafter \@secondoftwo
 \fi
}%
\providecommand \natexlab [1]{#1}%
\providecommand \enquote  [1]{``#1''}%
\providecommand \bibnamefont  [1]{#1}%
\providecommand \bibfnamefont [1]{#1}%
\providecommand \citenamefont [1]{#1}%
\providecommand \href@noop [0]{\@secondoftwo}%
\providecommand \href [0]{\begingroup \@sanitize@url \@href}%
\providecommand \@href[1]{\@@startlink{#1}\@@href}%
\providecommand \@@href[1]{\endgroup#1\@@endlink}%
\providecommand \@sanitize@url [0]{\catcode `\\12\catcode `\$12\catcode
  `\&12\catcode `\#12\catcode `\^12\catcode `\_12\catcode `\%12\relax}%
\providecommand \@@startlink[1]{}%
\providecommand \@@endlink[0]{}%
\providecommand \url  [0]{\begingroup\@sanitize@url \@url }%
\providecommand \@url [1]{\endgroup\@href {#1}{\urlprefix }}%
\providecommand \urlprefix  [0]{URL }%
\providecommand \Eprint [0]{\href }%
\providecommand \doibase [0]{http://dx.doi.org/}%
\providecommand \selectlanguage [0]{\@gobble}%
\providecommand \bibinfo  [0]{\@secondoftwo}%
\providecommand \bibfield  [0]{\@secondoftwo}%
\providecommand \translation [1]{[#1]}%
\providecommand \BibitemOpen [0]{}%
\providecommand \bibitemStop [0]{}%
\providecommand \bibitemNoStop [0]{.\EOS\space}%
\providecommand \EOS [0]{\spacefactor3000\relax}%
\providecommand \BibitemShut  [1]{\csname bibitem#1\endcsname}%
\let\auto@bib@innerbib\@empty
\bibitem [{\citenamefont {Randeria}\ \emph {et~al.}(1989)\citenamefont
  {Randeria}, \citenamefont {Duan},\ and\ \citenamefont
  {Shieh}}]{Randeria1989}%
  \BibitemOpen
  \bibfield  {author} {\bibinfo {author} {\bibfnamefont {M.}~\bibnamefont
  {Randeria}}, \bibinfo {author} {\bibfnamefont {J.-M.}\ \bibnamefont {Duan}},
  \ and\ \bibinfo {author} {\bibfnamefont {L.-Y.}\ \bibnamefont {Shieh}},\
  }\href {\doibase 10.1103/PhysRevLett.62.981} {\bibfield  {journal} {\bibinfo
  {journal} {Phys. Rev. Lett.}\ }\textbf {\bibinfo {volume} {62}},\ \bibinfo
  {pages} {981} (\bibinfo {year} {1989})}\BibitemShut {NoStop}%
\bibitem [{\citenamefont {Brodsky}\ \emph {et~al.}(2006)\citenamefont
  {Brodsky}, \citenamefont {Kagan}, \citenamefont {Klaptsov}, \citenamefont
  {Combescot},\ and\ \citenamefont {Leyronas}}]{Brodsky2006}%
  \BibitemOpen
  \bibfield  {author} {\bibinfo {author} {\bibfnamefont {I.~V.}\ \bibnamefont
  {Brodsky}}, \bibinfo {author} {\bibfnamefont {M.~Y.}\ \bibnamefont {Kagan}},
  \bibinfo {author} {\bibfnamefont {A.~V.}\ \bibnamefont {Klaptsov}}, \bibinfo
  {author} {\bibfnamefont {R.}~\bibnamefont {Combescot}}, \ and\ \bibinfo
  {author} {\bibfnamefont {X.}~\bibnamefont {Leyronas}},\ }\href {\doibase
  10.1103/PhysRevA.73.032724} {\bibfield  {journal} {\bibinfo  {journal} {Phys.
  Rev. A}\ }\textbf {\bibinfo {volume} {73}},\ \bibinfo {pages} {032724}
  (\bibinfo {year} {2006})}\BibitemShut {NoStop}%
\bibitem [{\citenamefont {Bertaina}\ and\ \citenamefont
  {Giorgini}(2011)}]{Bertaina2011}%
  \BibitemOpen
  \bibfield  {author} {\bibinfo {author} {\bibfnamefont {G.}~\bibnamefont
  {Bertaina}}\ and\ \bibinfo {author} {\bibfnamefont {S.}~\bibnamefont
  {Giorgini}},\ }\href {\doibase 10.1103/PhysRevLett.106.110403} {\bibfield
  {journal} {\bibinfo  {journal} {Phys. Rev. Lett.}\ }\textbf {\bibinfo
  {volume} {106}},\ \bibinfo {pages} {110403} (\bibinfo {year}
  {2011})}\BibitemShut {NoStop}%
\bibitem [{\citenamefont {Levinsen}\ and\ \citenamefont
  {Parish}(2015)}]{Levinsen2015}%
  \BibitemOpen
  \bibfield  {author} {\bibinfo {author} {\bibfnamefont {J.}~\bibnamefont
  {Levinsen}}\ and\ \bibinfo {author} {\bibfnamefont {M.~M.}\ \bibnamefont
  {Parish}},\ }\enquote {\bibinfo {title} {Strongly interacting two-dimensional
  fermi gases},}\ in\ \href {\doibase 10.1142/9789814667746_0001} {\emph
  {\bibinfo {booktitle} {Annual Review of Cold Atoms and Molecules}}}\
  (\bibinfo  {publisher} {World Scientific},\ \bibinfo {year} {2015})\
  Chap.~\bibinfo {chapter} {1}, pp.\ \bibinfo {pages} {1--75}\BibitemShut
  {NoStop}%
\bibitem [{\citenamefont {Berezinskii}(1971)}]{Berezinskii1971}%
  \BibitemOpen
  \bibfield  {author} {\bibinfo {author} {\bibfnamefont {V.~L.}\ \bibnamefont
  {Berezinskii}},\ }\href@noop {} {\bibfield  {journal} {\bibinfo  {journal}
  {Sov. Phys. JETP}\ }\textbf {\bibinfo {volume} {32}},\ \bibinfo {pages} {493}
  (\bibinfo {year} {1971})}\BibitemShut {NoStop}%
\bibitem [{\citenamefont {Berezinskii}(1972)}]{Berezinskii1972}%
  \BibitemOpen
  \bibfield  {author} {\bibinfo {author} {\bibfnamefont {V.~L.}\ \bibnamefont
  {Berezinskii}},\ }\href@noop {} {\bibfield  {journal} {\bibinfo  {journal}
  {Sov. Phys. JETP}\ }\textbf {\bibinfo {volume} {34}},\ \bibinfo {pages} {610}
  (\bibinfo {year} {1972})}\BibitemShut {NoStop}%
\bibitem [{\citenamefont {Kosterlitz}\ and\ \citenamefont
  {Thouless}(1973)}]{Kosterlitz1973}%
  \BibitemOpen
  \bibfield  {author} {\bibinfo {author} {\bibfnamefont {J.~M.}\ \bibnamefont
  {Kosterlitz}}\ and\ \bibinfo {author} {\bibfnamefont {D.~J.}\ \bibnamefont
  {Thouless}},\ }\href {\doibase 10.1088/0022-3719/6/7/010} {\bibfield
  {journal} {\bibinfo  {journal} {Journal of Physics C: Solid State Physics}\
  }\textbf {\bibinfo {volume} {6}},\ \bibinfo {pages} {1181} (\bibinfo {year}
  {1973})}\BibitemShut {NoStop}%
\bibitem [{\citenamefont {Fr\"ohlich}\ \emph {et~al.}(2011)\citenamefont
  {Fr\"ohlich}, \citenamefont {Feld}, \citenamefont {Vogt}, \citenamefont
  {Koschorreck}, \citenamefont {Zwerger},\ and\ \citenamefont
  {K\"ohl}}]{Frohlich2011}%
  \BibitemOpen
  \bibfield  {author} {\bibinfo {author} {\bibfnamefont {B.}~\bibnamefont
  {Fr\"ohlich}}, \bibinfo {author} {\bibfnamefont {M.}~\bibnamefont {Feld}},
  \bibinfo {author} {\bibfnamefont {E.}~\bibnamefont {Vogt}}, \bibinfo {author}
  {\bibfnamefont {M.}~\bibnamefont {Koschorreck}}, \bibinfo {author}
  {\bibfnamefont {W.}~\bibnamefont {Zwerger}}, \ and\ \bibinfo {author}
  {\bibfnamefont {M.}~\bibnamefont {K\"ohl}},\ }\href {\doibase
  10.1103/PhysRevLett.106.105301} {\bibfield  {journal} {\bibinfo  {journal}
  {Phys. Rev. Lett.}\ }\textbf {\bibinfo {volume} {106}},\ \bibinfo {pages}
  {105301} (\bibinfo {year} {2011})}\BibitemShut {NoStop}%
\bibitem [{\citenamefont {Vogt}\ \emph {et~al.}(2012)\citenamefont {Vogt},
  \citenamefont {Feld}, \citenamefont {Fr\"ohlich}, \citenamefont {Pertot},
  \citenamefont {Koschorreck},\ and\ \citenamefont {K\"ohl}}]{Vogt2012}%
  \BibitemOpen
  \bibfield  {author} {\bibinfo {author} {\bibfnamefont {E.}~\bibnamefont
  {Vogt}}, \bibinfo {author} {\bibfnamefont {M.}~\bibnamefont {Feld}}, \bibinfo
  {author} {\bibfnamefont {B.}~\bibnamefont {Fr\"ohlich}}, \bibinfo {author}
  {\bibfnamefont {D.}~\bibnamefont {Pertot}}, \bibinfo {author} {\bibfnamefont
  {M.}~\bibnamefont {Koschorreck}}, \ and\ \bibinfo {author} {\bibfnamefont
  {M.}~\bibnamefont {K\"ohl}},\ }\href {\doibase
  10.1103/PhysRevLett.108.070404} {\bibfield  {journal} {\bibinfo  {journal}
  {Phys. Rev. Lett.}\ }\textbf {\bibinfo {volume} {108}},\ \bibinfo {pages}
  {070404} (\bibinfo {year} {2012})}\BibitemShut {NoStop}%
\bibitem [{\citenamefont {Murthy}\ \emph {et~al.}(2015)\citenamefont {Murthy},
  \citenamefont {Boettcher}, \citenamefont {Bayha}, \citenamefont {Holzmann},
  \citenamefont {Kedar}, \citenamefont {Neidig}, \citenamefont {Ries},
  \citenamefont {Wenz}, \citenamefont {Z\"urn},\ and\ \citenamefont
  {Jochim}}]{Murthy2015}%
  \BibitemOpen
  \bibfield  {author} {\bibinfo {author} {\bibfnamefont {P.~A.}\ \bibnamefont
  {Murthy}}, \bibinfo {author} {\bibfnamefont {I.}~\bibnamefont {Boettcher}},
  \bibinfo {author} {\bibfnamefont {L.}~\bibnamefont {Bayha}}, \bibinfo
  {author} {\bibfnamefont {M.}~\bibnamefont {Holzmann}}, \bibinfo {author}
  {\bibfnamefont {D.}~\bibnamefont {Kedar}}, \bibinfo {author} {\bibfnamefont
  {M.}~\bibnamefont {Neidig}}, \bibinfo {author} {\bibfnamefont {M.~G.}\
  \bibnamefont {Ries}}, \bibinfo {author} {\bibfnamefont {A.~N.}\ \bibnamefont
  {Wenz}}, \bibinfo {author} {\bibfnamefont {G.}~\bibnamefont {Z\"urn}}, \ and\
  \bibinfo {author} {\bibfnamefont {S.}~\bibnamefont {Jochim}},\ }\href
  {\doibase 10.1103/PhysRevLett.115.010401} {\bibfield  {journal} {\bibinfo
  {journal} {Phys. Rev. Lett.}\ }\textbf {\bibinfo {volume} {115}},\ \bibinfo
  {pages} {010401} (\bibinfo {year} {2015})}\BibitemShut {NoStop}%
\bibitem [{\citenamefont {Ries}\ \emph {et~al.}(2015)\citenamefont {Ries},
  \citenamefont {Wenz}, \citenamefont {Z\"urn}, \citenamefont {Bayha},
  \citenamefont {Boettcher}, \citenamefont {Kedar}, \citenamefont {Murthy},
  \citenamefont {Neidig}, \citenamefont {Lompe},\ and\ \citenamefont
  {Jochim}}]{Ries2015}%
  \BibitemOpen
  \bibfield  {author} {\bibinfo {author} {\bibfnamefont {M.~G.}\ \bibnamefont
  {Ries}}, \bibinfo {author} {\bibfnamefont {A.~N.}\ \bibnamefont {Wenz}},
  \bibinfo {author} {\bibfnamefont {G.}~\bibnamefont {Z\"urn}}, \bibinfo
  {author} {\bibfnamefont {L.}~\bibnamefont {Bayha}}, \bibinfo {author}
  {\bibfnamefont {I.}~\bibnamefont {Boettcher}}, \bibinfo {author}
  {\bibfnamefont {D.}~\bibnamefont {Kedar}}, \bibinfo {author} {\bibfnamefont
  {P.~A.}\ \bibnamefont {Murthy}}, \bibinfo {author} {\bibfnamefont
  {M.}~\bibnamefont {Neidig}}, \bibinfo {author} {\bibfnamefont
  {T.}~\bibnamefont {Lompe}}, \ and\ \bibinfo {author} {\bibfnamefont
  {S.}~\bibnamefont {Jochim}},\ }\href {\doibase
  10.1103/PhysRevLett.114.230401} {\bibfield  {journal} {\bibinfo  {journal}
  {Phys. Rev. Lett.}\ }\textbf {\bibinfo {volume} {114}},\ \bibinfo {pages}
  {230401} (\bibinfo {year} {2015})}\BibitemShut {NoStop}%
\bibitem [{\citenamefont {Boettcher}\ \emph {et~al.}(2016)\citenamefont
  {Boettcher}, \citenamefont {Bayha}, \citenamefont {Kedar}, \citenamefont
  {Murthy}, \citenamefont {Neidig}, \citenamefont {Ries}, \citenamefont {Wenz},
  \citenamefont {Z\"urn}, \citenamefont {Jochim},\ and\ \citenamefont
  {Enss}}]{Boettcher2016}%
  \BibitemOpen
  \bibfield  {author} {\bibinfo {author} {\bibfnamefont {I.}~\bibnamefont
  {Boettcher}}, \bibinfo {author} {\bibfnamefont {L.}~\bibnamefont {Bayha}},
  \bibinfo {author} {\bibfnamefont {D.}~\bibnamefont {Kedar}}, \bibinfo
  {author} {\bibfnamefont {P.~A.}\ \bibnamefont {Murthy}}, \bibinfo {author}
  {\bibfnamefont {M.}~\bibnamefont {Neidig}}, \bibinfo {author} {\bibfnamefont
  {M.~G.}\ \bibnamefont {Ries}}, \bibinfo {author} {\bibfnamefont {A.~N.}\
  \bibnamefont {Wenz}}, \bibinfo {author} {\bibfnamefont {G.}~\bibnamefont
  {Z\"urn}}, \bibinfo {author} {\bibfnamefont {S.}~\bibnamefont {Jochim}}, \
  and\ \bibinfo {author} {\bibfnamefont {T.}~\bibnamefont {Enss}},\ }\href
  {\doibase 10.1103/PhysRevLett.116.045303} {\bibfield  {journal} {\bibinfo
  {journal} {Phys. Rev. Lett.}\ }\textbf {\bibinfo {volume} {116}},\ \bibinfo
  {pages} {045303} (\bibinfo {year} {2016})}\BibitemShut {NoStop}%
\bibitem [{\citenamefont {Fenech}\ \emph {et~al.}(2016)\citenamefont {Fenech},
  \citenamefont {Dyke}, \citenamefont {Peppler}, \citenamefont {Lingham},
  \citenamefont {Hoinka}, \citenamefont {Hu},\ and\ \citenamefont
  {Vale}}]{Fenech2016}%
  \BibitemOpen
  \bibfield  {author} {\bibinfo {author} {\bibfnamefont {K.}~\bibnamefont
  {Fenech}}, \bibinfo {author} {\bibfnamefont {P.}~\bibnamefont {Dyke}},
  \bibinfo {author} {\bibfnamefont {T.}~\bibnamefont {Peppler}}, \bibinfo
  {author} {\bibfnamefont {M.~G.}\ \bibnamefont {Lingham}}, \bibinfo {author}
  {\bibfnamefont {S.}~\bibnamefont {Hoinka}}, \bibinfo {author} {\bibfnamefont
  {H.}~\bibnamefont {Hu}}, \ and\ \bibinfo {author} {\bibfnamefont {C.~J.}\
  \bibnamefont {Vale}},\ }\href {\doibase 10.1103/PhysRevLett.116.045302}
  {\bibfield  {journal} {\bibinfo  {journal} {Phys. Rev. Lett.}\ }\textbf
  {\bibinfo {volume} {116}},\ \bibinfo {pages} {045302} (\bibinfo {year}
  {2016})}\BibitemShut {NoStop}%
\bibitem [{\citenamefont {Toniolo}\ \emph {et~al.}(2017)\citenamefont
  {Toniolo}, \citenamefont {Mulkerin}, \citenamefont {Vale}, \citenamefont
  {Liu},\ and\ \citenamefont {Hu}}]{Toniolo2017}%
  \BibitemOpen
  \bibfield  {author} {\bibinfo {author} {\bibfnamefont {U.}~\bibnamefont
  {Toniolo}}, \bibinfo {author} {\bibfnamefont {B.~C.}\ \bibnamefont
  {Mulkerin}}, \bibinfo {author} {\bibfnamefont {C.~J.}\ \bibnamefont {Vale}},
  \bibinfo {author} {\bibfnamefont {X.-J.}\ \bibnamefont {Liu}}, \ and\
  \bibinfo {author} {\bibfnamefont {H.}~\bibnamefont {Hu}},\ }\href {\doibase
  10.1103/PhysRevA.96.041604} {\bibfield  {journal} {\bibinfo  {journal} {Phys.
  Rev. A}\ }\textbf {\bibinfo {volume} {96}},\ \bibinfo {pages} {041604}
  (\bibinfo {year} {2017})}\BibitemShut {NoStop}%
\bibitem [{\citenamefont {Luciuk}\ \emph {et~al.}(2017)\citenamefont {Luciuk},
  \citenamefont {Smale}, \citenamefont {B\"ottcher}, \citenamefont {Sharum},
  \citenamefont {Olsen}, \citenamefont {Trotzky}, \citenamefont {Enss},\ and\
  \citenamefont {Thywissen}}]{Luciuk2017}%
  \BibitemOpen
  \bibfield  {author} {\bibinfo {author} {\bibfnamefont {C.}~\bibnamefont
  {Luciuk}}, \bibinfo {author} {\bibfnamefont {S.}~\bibnamefont {Smale}},
  \bibinfo {author} {\bibfnamefont {F.}~\bibnamefont {B\"ottcher}}, \bibinfo
  {author} {\bibfnamefont {H.}~\bibnamefont {Sharum}}, \bibinfo {author}
  {\bibfnamefont {B.~A.}\ \bibnamefont {Olsen}}, \bibinfo {author}
  {\bibfnamefont {S.}~\bibnamefont {Trotzky}}, \bibinfo {author} {\bibfnamefont
  {T.}~\bibnamefont {Enss}}, \ and\ \bibinfo {author} {\bibfnamefont {J.~H.}\
  \bibnamefont {Thywissen}},\ }\href {\doibase 10.1103/PhysRevLett.118.130405}
  {\bibfield  {journal} {\bibinfo  {journal} {Phys. Rev. Lett.}\ }\textbf
  {\bibinfo {volume} {118}},\ \bibinfo {pages} {130405} (\bibinfo {year}
  {2017})}\BibitemShut {NoStop}%
\bibitem [{\citenamefont {Hueck}\ \emph {et~al.}(2018)\citenamefont {Hueck},
  \citenamefont {Luick}, \citenamefont {Sobirey}, \citenamefont {Siegl},
  \citenamefont {Lompe},\ and\ \citenamefont {Moritz}}]{Hueck2018}%
  \BibitemOpen
  \bibfield  {author} {\bibinfo {author} {\bibfnamefont {K.}~\bibnamefont
  {Hueck}}, \bibinfo {author} {\bibfnamefont {N.}~\bibnamefont {Luick}},
  \bibinfo {author} {\bibfnamefont {L.}~\bibnamefont {Sobirey}}, \bibinfo
  {author} {\bibfnamefont {J.}~\bibnamefont {Siegl}}, \bibinfo {author}
  {\bibfnamefont {T.}~\bibnamefont {Lompe}}, \ and\ \bibinfo {author}
  {\bibfnamefont {H.}~\bibnamefont {Moritz}},\ }\href {\doibase
  10.1103/PhysRevLett.120.060402} {\bibfield  {journal} {\bibinfo  {journal}
  {Phys. Rev. Lett.}\ }\textbf {\bibinfo {volume} {120}},\ \bibinfo {pages}
  {060402} (\bibinfo {year} {2018})}\BibitemShut {NoStop}%
\bibitem [{\citenamefont {Murthy}\ \emph {et~al.}(2018)\citenamefont {Murthy},
  \citenamefont {Neidig}, \citenamefont {Klemt}, \citenamefont {Bayha},
  \citenamefont {Boettcher}, \citenamefont {Enss}, \citenamefont {Holten},
  \citenamefont {Z{\"u}rn}, \citenamefont {Preiss},\ and\ \citenamefont
  {Jochim}}]{Murthy2018}%
  \BibitemOpen
  \bibfield  {author} {\bibinfo {author} {\bibfnamefont {P.~A.}\ \bibnamefont
  {Murthy}}, \bibinfo {author} {\bibfnamefont {M.}~\bibnamefont {Neidig}},
  \bibinfo {author} {\bibfnamefont {R.}~\bibnamefont {Klemt}}, \bibinfo
  {author} {\bibfnamefont {L.}~\bibnamefont {Bayha}}, \bibinfo {author}
  {\bibfnamefont {I.}~\bibnamefont {Boettcher}}, \bibinfo {author}
  {\bibfnamefont {T.}~\bibnamefont {Enss}}, \bibinfo {author} {\bibfnamefont
  {M.}~\bibnamefont {Holten}}, \bibinfo {author} {\bibfnamefont
  {G.}~\bibnamefont {Z{\"u}rn}}, \bibinfo {author} {\bibfnamefont {P.~M.}\
  \bibnamefont {Preiss}}, \ and\ \bibinfo {author} {\bibfnamefont
  {S.}~\bibnamefont {Jochim}},\ }\href {\doibase 10.1126/science.aan5950}
  {\bibfield  {journal} {\bibinfo  {journal} {Science}\ }\textbf {\bibinfo
  {volume} {359}},\ \bibinfo {pages} {452} (\bibinfo {year}
  {2018})}\BibitemShut {NoStop}%
\bibitem [{\citenamefont {Watanabe}\ \emph {et~al.}(2013)\citenamefont
  {Watanabe}, \citenamefont {Tsuchiya},\ and\ \citenamefont
  {Ohashi}}]{Watanabe2013}%
  \BibitemOpen
  \bibfield  {author} {\bibinfo {author} {\bibfnamefont {R.}~\bibnamefont
  {Watanabe}}, \bibinfo {author} {\bibfnamefont {S.}~\bibnamefont {Tsuchiya}},
  \ and\ \bibinfo {author} {\bibfnamefont {Y.}~\bibnamefont {Ohashi}},\ }\href
  {\doibase 10.1007/s10909-012-0691-7} {\bibfield  {journal} {\bibinfo
  {journal} {Journal of Low Temperature Physics}\ }\textbf {\bibinfo {volume}
  {171}},\ \bibinfo {pages} {341} (\bibinfo {year} {2013})}\BibitemShut
  {NoStop}%
\bibitem [{\citenamefont {Matsumoto}\ and\ \citenamefont
  {Ohashi}(2014)}]{Matsumoto2014}%
  \BibitemOpen
  \bibfield  {author} {\bibinfo {author} {\bibfnamefont {M.}~\bibnamefont
  {Matsumoto}}\ and\ \bibinfo {author} {\bibfnamefont {Y.}~\bibnamefont
  {Ohashi}},\ }\href {\doibase 10.1088/1742-6596/568/1/012012} {\bibfield
  {journal} {\bibinfo  {journal} {Journal of Physics: Conference Series}\
  }\textbf {\bibinfo {volume} {568}},\ \bibinfo {pages} {012012} (\bibinfo
  {year} {2014})}\BibitemShut {NoStop}%
\bibitem [{\citenamefont {Bauer}\ \emph {et~al.}(2014)\citenamefont {Bauer},
  \citenamefont {Parish},\ and\ \citenamefont {Enss}}]{Bauer2014}%
  \BibitemOpen
  \bibfield  {author} {\bibinfo {author} {\bibfnamefont {M.}~\bibnamefont
  {Bauer}}, \bibinfo {author} {\bibfnamefont {M.~M.}\ \bibnamefont {Parish}}, \
  and\ \bibinfo {author} {\bibfnamefont {T.}~\bibnamefont {Enss}},\ }\href
  {\doibase 10.1103/PhysRevLett.112.135302} {\bibfield  {journal} {\bibinfo
  {journal} {Phys. Rev. Lett.}\ }\textbf {\bibinfo {volume} {112}},\ \bibinfo
  {pages} {135302} (\bibinfo {year} {2014})}\BibitemShut {NoStop}%
\bibitem [{\citenamefont {Anderson}\ and\ \citenamefont
  {Drut}(2015)}]{Anderson2015}%
  \BibitemOpen
  \bibfield  {author} {\bibinfo {author} {\bibfnamefont {E.~R.}\ \bibnamefont
  {Anderson}}\ and\ \bibinfo {author} {\bibfnamefont {J.~E.}\ \bibnamefont
  {Drut}},\ }\href {\doibase 10.1103/PhysRevLett.115.115301} {\bibfield
  {journal} {\bibinfo  {journal} {Phys. Rev. Lett.}\ }\textbf {\bibinfo
  {volume} {115}},\ \bibinfo {pages} {115301} (\bibinfo {year}
  {2015})}\BibitemShut {NoStop}%
\bibitem [{\citenamefont {Marsiglio}\ \emph {et~al.}(2015)\citenamefont
  {Marsiglio}, \citenamefont {Pieri}, \citenamefont {Perali}, \citenamefont
  {Palestini},\ and\ \citenamefont {Strinati}}]{Marsiglio2015}%
  \BibitemOpen
  \bibfield  {author} {\bibinfo {author} {\bibfnamefont {F.}~\bibnamefont
  {Marsiglio}}, \bibinfo {author} {\bibfnamefont {P.}~\bibnamefont {Pieri}},
  \bibinfo {author} {\bibfnamefont {A.}~\bibnamefont {Perali}}, \bibinfo
  {author} {\bibfnamefont {F.}~\bibnamefont {Palestini}}, \ and\ \bibinfo
  {author} {\bibfnamefont {G.~C.}\ \bibnamefont {Strinati}},\ }\href {\doibase
  10.1103/PhysRevB.91.054509} {\bibfield  {journal} {\bibinfo  {journal} {Phys.
  Rev. B}\ }\textbf {\bibinfo {volume} {91}},\ \bibinfo {pages} {054509}
  (\bibinfo {year} {2015})}\BibitemShut {NoStop}%
\bibitem [{\citenamefont {Shi}\ \emph {et~al.}(2015)\citenamefont {Shi},
  \citenamefont {Chiesa},\ and\ \citenamefont {Zhang}}]{Shi2015}%
  \BibitemOpen
  \bibfield  {author} {\bibinfo {author} {\bibfnamefont {H.}~\bibnamefont
  {Shi}}, \bibinfo {author} {\bibfnamefont {S.}~\bibnamefont {Chiesa}}, \ and\
  \bibinfo {author} {\bibfnamefont {S.}~\bibnamefont {Zhang}},\ }\href
  {\doibase 10.1103/PhysRevA.92.033603} {\bibfield  {journal} {\bibinfo
  {journal} {Phys. Rev. A}\ }\textbf {\bibinfo {volume} {92}},\ \bibinfo
  {pages} {033603} (\bibinfo {year} {2015})}\BibitemShut {NoStop}%
\bibitem [{\citenamefont {Galea}\ \emph {et~al.}(2016)\citenamefont {Galea},
  \citenamefont {Dawkins}, \citenamefont {Gandolfi},\ and\ \citenamefont
  {Gezerlis}}]{Galea2016}%
  \BibitemOpen
  \bibfield  {author} {\bibinfo {author} {\bibfnamefont {A.}~\bibnamefont
  {Galea}}, \bibinfo {author} {\bibfnamefont {H.}~\bibnamefont {Dawkins}},
  \bibinfo {author} {\bibfnamefont {S.}~\bibnamefont {Gandolfi}}, \ and\
  \bibinfo {author} {\bibfnamefont {A.}~\bibnamefont {Gezerlis}},\ }\href
  {\doibase 10.1103/PhysRevA.93.023602} {\bibfield  {journal} {\bibinfo
  {journal} {Phys. Rev. A}\ }\textbf {\bibinfo {volume} {93}},\ \bibinfo
  {pages} {023602} (\bibinfo {year} {2016})}\BibitemShut {NoStop}%
\bibitem [{\citenamefont {Vitali}\ \emph {et~al.}(2017)\citenamefont {Vitali},
  \citenamefont {Shi}, \citenamefont {Qin},\ and\ \citenamefont
  {Zhang}}]{Vitali2017}%
  \BibitemOpen
  \bibfield  {author} {\bibinfo {author} {\bibfnamefont {E.}~\bibnamefont
  {Vitali}}, \bibinfo {author} {\bibfnamefont {H.}~\bibnamefont {Shi}},
  \bibinfo {author} {\bibfnamefont {M.}~\bibnamefont {Qin}}, \ and\ \bibinfo
  {author} {\bibfnamefont {S.}~\bibnamefont {Zhang}},\ }\href {\doibase
  10.1103/PhysRevA.96.061601} {\bibfield  {journal} {\bibinfo  {journal} {Phys.
  Rev. A}\ }\textbf {\bibinfo {volume} {96}},\ \bibinfo {pages} {061601}
  (\bibinfo {year} {2017})}\BibitemShut {NoStop}%
\bibitem [{\citenamefont {Madeira}\ \emph {et~al.}(2017)\citenamefont
  {Madeira}, \citenamefont {Gandolfi},\ and\ \citenamefont
  {Schmidt}}]{Madeira2017}%
  \BibitemOpen
  \bibfield  {author} {\bibinfo {author} {\bibfnamefont {L.}~\bibnamefont
  {Madeira}}, \bibinfo {author} {\bibfnamefont {S.}~\bibnamefont {Gandolfi}}, \
  and\ \bibinfo {author} {\bibfnamefont {K.~E.}\ \bibnamefont {Schmidt}},\
  }\href {\doibase 10.1103/PhysRevA.95.053603} {\bibfield  {journal} {\bibinfo
  {journal} {Phys. Rev. A}\ }\textbf {\bibinfo {volume} {95}},\ \bibinfo
  {pages} {053603} (\bibinfo {year} {2017})}\BibitemShut {NoStop}%
\bibitem [{\citenamefont {Schonenberg}\ \emph {et~al.}(2017)\citenamefont
  {Schonenberg}, \citenamefont {Verpoort},\ and\ \citenamefont
  {Conduit}}]{Schonenberg2017}%
  \BibitemOpen
  \bibfield  {author} {\bibinfo {author} {\bibfnamefont {L.~M.}\ \bibnamefont
  {Schonenberg}}, \bibinfo {author} {\bibfnamefont {P.~C.}\ \bibnamefont
  {Verpoort}}, \ and\ \bibinfo {author} {\bibfnamefont {G.~J.}\ \bibnamefont
  {Conduit}},\ }\href {\doibase 10.1103/PhysRevA.96.023619} {\bibfield
  {journal} {\bibinfo  {journal} {Phys. Rev. A}\ }\textbf {\bibinfo {volume}
  {96}},\ \bibinfo {pages} {023619} (\bibinfo {year} {2017})}\BibitemShut
  {NoStop}%
\bibitem [{\citenamefont {Mulkerin}\ \emph {et~al.}(2018)\citenamefont
  {Mulkerin}, \citenamefont {Liu},\ and\ \citenamefont {Hu}}]{Mulkerin2018}%
  \BibitemOpen
  \bibfield  {author} {\bibinfo {author} {\bibfnamefont {B.~C.}\ \bibnamefont
  {Mulkerin}}, \bibinfo {author} {\bibfnamefont {X.-J.}\ \bibnamefont {Liu}}, \
  and\ \bibinfo {author} {\bibfnamefont {H.}~\bibnamefont {Hu}},\ }\href
  {\doibase 10.1103/PhysRevA.97.053612} {\bibfield  {journal} {\bibinfo
  {journal} {Phys. Rev. A}\ }\textbf {\bibinfo {volume} {97}},\ \bibinfo
  {pages} {053612} (\bibinfo {year} {2018})}\BibitemShut {NoStop}%
\bibitem [{\citenamefont {Hu}\ \emph {et~al.}(2019)\citenamefont {Hu},
  \citenamefont {Mulkerin}, \citenamefont {Toniolo}, \citenamefont {He},\ and\
  \citenamefont {Liu}}]{Hu2019}%
  \BibitemOpen
  \bibfield  {author} {\bibinfo {author} {\bibfnamefont {H.}~\bibnamefont
  {Hu}}, \bibinfo {author} {\bibfnamefont {B.~C.}\ \bibnamefont {Mulkerin}},
  \bibinfo {author} {\bibfnamefont {U.}~\bibnamefont {Toniolo}}, \bibinfo
  {author} {\bibfnamefont {L.}~\bibnamefont {He}}, \ and\ \bibinfo {author}
  {\bibfnamefont {X.-J.}\ \bibnamefont {Liu}},\ }\href {\doibase
  10.1103/PhysRevLett.122.070401} {\bibfield  {journal} {\bibinfo  {journal}
  {Phys. Rev. Lett.}\ }\textbf {\bibinfo {volume} {122}},\ \bibinfo {pages}
  {070401} (\bibinfo {year} {2019})}\BibitemShut {NoStop}%
\bibitem [{\citenamefont {Wu}\ \emph {et~al.}(2020)\citenamefont {Wu},
  \citenamefont {Hu}, \citenamefont {He}, \citenamefont {Liu},\ and\
  \citenamefont {Hu}}]{Wu2020}%
  \BibitemOpen
  \bibfield  {author} {\bibinfo {author} {\bibfnamefont {F.}~\bibnamefont
  {Wu}}, \bibinfo {author} {\bibfnamefont {J.}~\bibnamefont {Hu}}, \bibinfo
  {author} {\bibfnamefont {L.}~\bibnamefont {He}}, \bibinfo {author}
  {\bibfnamefont {X.-J.}\ \bibnamefont {Liu}}, \ and\ \bibinfo {author}
  {\bibfnamefont {H.}~\bibnamefont {Hu}},\ }\href {\doibase
  10.1103/PhysRevA.101.043607} {\bibfield  {journal} {\bibinfo  {journal}
  {Phys. Rev. A}\ }\textbf {\bibinfo {volume} {101}},\ \bibinfo {pages}
  {043607} (\bibinfo {year} {2020})}\BibitemShut {NoStop}%
\bibitem [{\citenamefont {Pascucci}\ and\ \citenamefont
  {Salasnich}(2020)}]{Pascucci2020}%
  \BibitemOpen
  \bibfield  {author} {\bibinfo {author} {\bibfnamefont {F.}~\bibnamefont
  {Pascucci}}\ and\ \bibinfo {author} {\bibfnamefont {L.}~\bibnamefont
  {Salasnich}},\ }\href {\doibase 10.1103/PhysRevA.102.013325} {\bibfield
  {journal} {\bibinfo  {journal} {Phys. Rev. A}\ }\textbf {\bibinfo {volume}
  {102}},\ \bibinfo {pages} {013325} (\bibinfo {year} {2020})}\BibitemShut
  {NoStop}%
\bibitem [{\citenamefont {Zhao}\ \emph {et~al.}(2020)\citenamefont {Zhao},
  \citenamefont {Gao}, \citenamefont {Liang}, \citenamefont {Zou},\ and\
  \citenamefont {Yuan}}]{Zhao2020}%
  \BibitemOpen
  \bibfield  {author} {\bibinfo {author} {\bibfnamefont {H.}~\bibnamefont
  {Zhao}}, \bibinfo {author} {\bibfnamefont {X.}~\bibnamefont {Gao}}, \bibinfo
  {author} {\bibfnamefont {W.}~\bibnamefont {Liang}}, \bibinfo {author}
  {\bibfnamefont {P.}~\bibnamefont {Zou}}, \ and\ \bibinfo {author}
  {\bibfnamefont {F.}~\bibnamefont {Yuan}},\ }\href {\doibase
  10.1088/1367-2630/abab3d} {\bibfield  {journal} {\bibinfo  {journal} {New
  Journal of Physics}\ }\textbf {\bibinfo {volume} {22}},\ \bibinfo {pages}
  {093012} (\bibinfo {year} {2020})}\BibitemShut {NoStop}%
\bibitem [{\citenamefont {Zielinski}\ \emph {et~al.}(2020)\citenamefont
  {Zielinski}, \citenamefont {Ross},\ and\ \citenamefont
  {Gezerlis}}]{Zielinski2020}%
  \BibitemOpen
  \bibfield  {author} {\bibinfo {author} {\bibfnamefont {T.}~\bibnamefont
  {Zielinski}}, \bibinfo {author} {\bibfnamefont {B.}~\bibnamefont {Ross}}, \
  and\ \bibinfo {author} {\bibfnamefont {A.}~\bibnamefont {Gezerlis}},\ }\href
  {\doibase 10.1103/PhysRevA.101.033601} {\bibfield  {journal} {\bibinfo
  {journal} {Phys. Rev. A}\ }\textbf {\bibinfo {volume} {101}},\ \bibinfo
  {pages} {033601} (\bibinfo {year} {2020})}\BibitemShut {NoStop}%
\bibitem [{\citenamefont {Mulkerin}\ \emph
  {et~al.}(2020{\natexlab{a}})\citenamefont {Mulkerin}, \citenamefont {Hu},\
  and\ \citenamefont {Liu}}]{Mulkerin2020a}%
  \BibitemOpen
  \bibfield  {author} {\bibinfo {author} {\bibfnamefont {B.~C.}\ \bibnamefont
  {Mulkerin}}, \bibinfo {author} {\bibfnamefont {H.}~\bibnamefont {Hu}}, \ and\
  \bibinfo {author} {\bibfnamefont {X.-J.}\ \bibnamefont {Liu}},\ }\href
  {\doibase 10.1103/PhysRevA.101.013605} {\bibfield  {journal} {\bibinfo
  {journal} {Phys. Rev. A}\ }\textbf {\bibinfo {volume} {101}},\ \bibinfo
  {pages} {013605} (\bibinfo {year} {2020}{\natexlab{a}})}\BibitemShut
  {NoStop}%
\bibitem [{\citenamefont {Mulkerin}\ \emph
  {et~al.}(2020{\natexlab{b}})\citenamefont {Mulkerin}, \citenamefont {Liu},\
  and\ \citenamefont {Hu}}]{Mulkerin2020b}%
  \BibitemOpen
  \bibfield  {author} {\bibinfo {author} {\bibfnamefont {B.~C.}\ \bibnamefont
  {Mulkerin}}, \bibinfo {author} {\bibfnamefont {X.-J.}\ \bibnamefont {Liu}}, \
  and\ \bibinfo {author} {\bibfnamefont {H.}~\bibnamefont {Hu}},\ }\href
  {\doibase 10.1103/PhysRevA.102.013313} {\bibfield  {journal} {\bibinfo
  {journal} {Phys. Rev. A}\ }\textbf {\bibinfo {volume} {102}},\ \bibinfo
  {pages} {013313} (\bibinfo {year} {2020}{\natexlab{b}})}\BibitemShut
  {NoStop}%
\bibitem [{\citenamefont {Wang}\ \emph {et~al.}(2020)\citenamefont {Wang},
  \citenamefont {Chen},\ and\ \citenamefont {Levin}}]{Wang2020}%
  \BibitemOpen
  \bibfield  {author} {\bibinfo {author} {\bibfnamefont {X.}~\bibnamefont
  {Wang}}, \bibinfo {author} {\bibfnamefont {Q.}~\bibnamefont {Chen}}, \ and\
  \bibinfo {author} {\bibfnamefont {K.}~\bibnamefont {Levin}},\ }\href
  {\doibase 10.1088/1367-2630/ab890b} {\bibfield  {journal} {\bibinfo
  {journal} {New Journal of Physics}\ }\textbf {\bibinfo {volume} {22}},\
  \bibinfo {pages} {063050} (\bibinfo {year} {2020})}\BibitemShut {NoStop}%
\bibitem [{\citenamefont {He}\ \emph {et~al.}(2022)\citenamefont {He},
  \citenamefont {Shi},\ and\ \citenamefont {Zhang}}]{He2022}%
  \BibitemOpen
  \bibfield  {author} {\bibinfo {author} {\bibfnamefont {Y.-Y.}\ \bibnamefont
  {He}}, \bibinfo {author} {\bibfnamefont {H.}~\bibnamefont {Shi}}, \ and\
  \bibinfo {author} {\bibfnamefont {S.}~\bibnamefont {Zhang}},\ }\href
  {\doibase 10.1103/PhysRevLett.129.076403} {\bibfield  {journal} {\bibinfo
  {journal} {Phys. Rev. Lett.}\ }\textbf {\bibinfo {volume} {129}},\ \bibinfo
  {pages} {076403} (\bibinfo {year} {2022})}\BibitemShut {NoStop}%
\bibitem [{\citenamefont {Feld}\ \emph {et~al.}(2011)\citenamefont {Feld},
  \citenamefont {Fr{\"o}hlich}, \citenamefont {Vogt}, \citenamefont
  {Koschorreck},\ and\ \citenamefont {K{\"o}hl}}]{Feld2011}%
  \BibitemOpen
  \bibfield  {author} {\bibinfo {author} {\bibfnamefont {M.}~\bibnamefont
  {Feld}}, \bibinfo {author} {\bibfnamefont {B.}~\bibnamefont {Fr{\"o}hlich}},
  \bibinfo {author} {\bibfnamefont {E.}~\bibnamefont {Vogt}}, \bibinfo {author}
  {\bibfnamefont {M.}~\bibnamefont {Koschorreck}}, \ and\ \bibinfo {author}
  {\bibfnamefont {M.}~\bibnamefont {K{\"o}hl}},\ }\href {\doibase
  10.1038/nature10627} {\bibfield  {journal} {\bibinfo  {journal} {Nature}\
  }\textbf {\bibinfo {volume} {480}},\ \bibinfo {pages} {75} (\bibinfo {year}
  {2011})}\BibitemShut {NoStop}%
\bibitem [{\citenamefont {Haussmann}\ \emph {et~al.}(2009)\citenamefont
  {Haussmann}, \citenamefont {Punk},\ and\ \citenamefont
  {Zwerger}}]{Haussmann2009}%
  \BibitemOpen
  \bibfield  {author} {\bibinfo {author} {\bibfnamefont {R.}~\bibnamefont
  {Haussmann}}, \bibinfo {author} {\bibfnamefont {M.}~\bibnamefont {Punk}}, \
  and\ \bibinfo {author} {\bibfnamefont {W.}~\bibnamefont {Zwerger}},\ }\href
  {\doibase 10.1103/PhysRevA.80.063612} {\bibfield  {journal} {\bibinfo
  {journal} {Phys. Rev. A}\ }\textbf {\bibinfo {volume} {80}},\ \bibinfo
  {pages} {063612} (\bibinfo {year} {2009})}\BibitemShut {NoStop}%
\bibitem [{\citenamefont {Zwerger}(2016)}]{Zwerger2016}%
  \BibitemOpen
  \bibfield  {author} {\bibinfo {author} {\bibfnamefont {W.}~\bibnamefont
  {Zwerger}},\ }\enquote {\bibinfo {title} {Strongly interacting fermi
  gases},}\ in\ \href@noop {} {\emph {\bibinfo {booktitle} {Proceedings of the
  International School of Physics \enquote{Enrico Fermi} - Course 191
  \enquote{Quantum Matter at Ultralow Temperatures}}}},\ \bibinfo {editor}
  {edited by\ \bibinfo {editor} {\bibfnamefont {M.}~\bibnamefont {Inguscio}},
  \bibinfo {editor} {\bibfnamefont {W.}~\bibnamefont {Ketterle}}, \bibinfo
  {editor} {\bibfnamefont {S.}~\bibnamefont {Stringari}}, \ and\ \bibinfo
  {editor} {\bibfnamefont {G.}~\bibnamefont {Roati}}}\ (\bibinfo  {publisher}
  {IOS Press},\ \bibinfo {address} {Amsterdam, SIF Bologna},\ \bibinfo {year}
  {2016})\ pp.\ \bibinfo {pages} {63--141}\BibitemShut {NoStop}%
\bibitem [{\citenamefont {Pini}\ \emph {et~al.}(2019)\citenamefont {Pini},
  \citenamefont {Pieri},\ and\ \citenamefont {Strinati}}]{Pini2019}%
  \BibitemOpen
  \bibfield  {author} {\bibinfo {author} {\bibfnamefont {M.}~\bibnamefont
  {Pini}}, \bibinfo {author} {\bibfnamefont {P.}~\bibnamefont {Pieri}}, \ and\
  \bibinfo {author} {\bibfnamefont {G.~C.}\ \bibnamefont {Strinati}},\ }\href
  {\doibase 10.1103/PhysRevB.99.094502} {\bibfield  {journal} {\bibinfo
  {journal} {Phys. Rev. B}\ }\textbf {\bibinfo {volume} {99}},\ \bibinfo
  {pages} {094502} (\bibinfo {year} {2019})}\BibitemShut {NoStop}%
\bibitem [{\citenamefont {Wlaz\l{}owski}\ \emph {et~al.}(2013)\citenamefont
  {Wlaz\l{}owski}, \citenamefont {Magierski}, \citenamefont {Drut},
  \citenamefont {Bulgac},\ and\ \citenamefont {Roche}}]{Wlaz?owski2013}%
  \BibitemOpen
  \bibfield  {author} {\bibinfo {author} {\bibfnamefont {G.}~\bibnamefont
  {Wlaz\l{}owski}}, \bibinfo {author} {\bibfnamefont {P.}~\bibnamefont
  {Magierski}}, \bibinfo {author} {\bibfnamefont {J.~E.}\ \bibnamefont {Drut}},
  \bibinfo {author} {\bibfnamefont {A.}~\bibnamefont {Bulgac}}, \ and\ \bibinfo
  {author} {\bibfnamefont {K.~J.}\ \bibnamefont {Roche}},\ }\href {\doibase
  10.1103/PhysRevLett.110.090401} {\bibfield  {journal} {\bibinfo  {journal}
  {Phys. Rev. Lett.}\ }\textbf {\bibinfo {volume} {110}},\ \bibinfo {pages}
  {090401} (\bibinfo {year} {2013})}\BibitemShut {NoStop}%
\bibitem [{\citenamefont {Jensen}\ \emph
  {et~al.}(2020{\natexlab{a}})\citenamefont {Jensen}, \citenamefont
  {Gilbreth},\ and\ \citenamefont {Alhassid}}]{Jensen2020a}%
  \BibitemOpen
  \bibfield  {author} {\bibinfo {author} {\bibfnamefont {S.}~\bibnamefont
  {Jensen}}, \bibinfo {author} {\bibfnamefont {C.~N.}\ \bibnamefont
  {Gilbreth}}, \ and\ \bibinfo {author} {\bibfnamefont {Y.}~\bibnamefont
  {Alhassid}},\ }\href {\doibase 10.1103/PhysRevLett.124.090604} {\bibfield
  {journal} {\bibinfo  {journal} {Phys. Rev. Lett.}\ }\textbf {\bibinfo
  {volume} {124}},\ \bibinfo {pages} {090604} (\bibinfo {year}
  {2020}{\natexlab{a}})}\BibitemShut {NoStop}%
\bibitem [{\citenamefont {Richie-Halford}\ \emph {et~al.}(2020)\citenamefont
  {Richie-Halford}, \citenamefont {Drut},\ and\ \citenamefont
  {Bulgac}}]{Richie2020}%
  \BibitemOpen
  \bibfield  {author} {\bibinfo {author} {\bibfnamefont {A.}~\bibnamefont
  {Richie-Halford}}, \bibinfo {author} {\bibfnamefont {J.~E.}\ \bibnamefont
  {Drut}}, \ and\ \bibinfo {author} {\bibfnamefont {A.}~\bibnamefont
  {Bulgac}},\ }\href {\doibase 10.1103/PhysRevLett.125.060403} {\bibfield
  {journal} {\bibinfo  {journal} {Phys. Rev. Lett.}\ }\textbf {\bibinfo
  {volume} {125}},\ \bibinfo {pages} {060403} (\bibinfo {year}
  {2020})}\BibitemShut {NoStop}%
\bibitem [{\citenamefont {Rammelm\"uller}\ \emph {et~al.}(2021)\citenamefont
  {Rammelm\"uller}, \citenamefont {Hou}, \citenamefont {Drut},\ and\
  \citenamefont {Braun}}]{Rammelmuller2021}%
  \BibitemOpen
  \bibfield  {author} {\bibinfo {author} {\bibfnamefont {L.}~\bibnamefont
  {Rammelm\"uller}}, \bibinfo {author} {\bibfnamefont {Y.}~\bibnamefont {Hou}},
  \bibinfo {author} {\bibfnamefont {J.~E.}\ \bibnamefont {Drut}}, \ and\
  \bibinfo {author} {\bibfnamefont {J.}~\bibnamefont {Braun}},\ }\href
  {\doibase 10.1103/PhysRevA.103.043330} {\bibfield  {journal} {\bibinfo
  {journal} {Phys. Rev. A}\ }\textbf {\bibinfo {volume} {103}},\ \bibinfo
  {pages} {043330} (\bibinfo {year} {2021})}\BibitemShut {NoStop}%
\bibitem [{\citenamefont {Alhassid}(2017)}]{Alhassid2017}%
  \BibitemOpen
  \bibfield  {author} {\bibinfo {author} {\bibfnamefont {Y.}~\bibnamefont
  {Alhassid}},\ }in\ \href@noop {} {\emph {\bibinfo {booktitle} {Emergent
  Phenomena in Atomic Nuclei from Large-Scale Modeling: a Symmetry-Guided
  Perspective}}},\ \bibinfo {editor} {edited by\ \bibinfo {editor}
  {\bibfnamefont {K.~D.}\ \bibnamefont {Launey}}}\ (\bibinfo  {publisher}
  {World Scientific},\ \bibinfo {year} {2017})\BibitemShut {NoStop}%
\bibitem [{\citenamefont {Jensen}\ \emph {et~al.}(2019)\citenamefont {Jensen},
  \citenamefont {Gilbreth},\ and\ \citenamefont {Alhassid}}]{Jensen2019}%
  \BibitemOpen
  \bibfield  {author} {\bibinfo {author} {\bibfnamefont {S.}~\bibnamefont
  {Jensen}}, \bibinfo {author} {\bibfnamefont {C.~N.}\ \bibnamefont
  {Gilbreth}}, \ and\ \bibinfo {author} {\bibfnamefont {Y.}~\bibnamefont
  {Alhassid}},\ }\href@noop {} {\bibfield  {journal} {\bibinfo  {journal}
  {European Journal of Physics: Special Topics}\ }\textbf {\bibinfo {volume}
  {227}},\ \bibinfo {pages} {2241} (\bibinfo {year} {2019})}\BibitemShut
  {NoStop}%
\bibitem [{\citenamefont {Alhassid}\ \emph {et~al.}(1999)\citenamefont
  {Alhassid}, \citenamefont {Liu},\ and\ \citenamefont
  {Nakada}}]{Alhassid1999}%
  \BibitemOpen
  \bibfield  {author} {\bibinfo {author} {\bibfnamefont {Y.}~\bibnamefont
  {Alhassid}}, \bibinfo {author} {\bibfnamefont {S.}~\bibnamefont {Liu}}, \
  and\ \bibinfo {author} {\bibfnamefont {H.}~\bibnamefont {Nakada}},\ }\href
  {\doibase 10.1103/PhysRevLett.83.4265} {\bibfield  {journal} {\bibinfo
  {journal} {Phys. Rev. Lett.}\ }\textbf {\bibinfo {volume} {83}},\ \bibinfo
  {pages} {4265} (\bibinfo {year} {1999})}\BibitemShut {NoStop}%
\bibitem [{\citenamefont {Jensen}\ \emph {et~al.}()\citenamefont {Jensen},
  \citenamefont {Gilbreth},\ and\ \citenamefont {Alhassid}}]{Jensen2023}%
  \BibitemOpen
  \bibfield  {author} {\bibinfo {author} {\bibfnamefont {S.}~\bibnamefont
  {Jensen}}, \bibinfo {author} {\bibfnamefont {C.~N.}\ \bibnamefont
  {Gilbreth}}, \ and\ \bibinfo {author} {\bibfnamefont {Y.}~\bibnamefont
  {Alhassid}},\ }\href@noop {} {\bibinfo  {journal} {to be published}\
  }\BibitemShut {NoStop}%
\bibitem [{\citenamefont {Tan}(2008)}]{Tan2008}%
  \BibitemOpen
\bibfield  {journal} {  }\bibfield  {author} {\bibinfo {author} {\bibfnamefont
  {S.}~\bibnamefont {Tan}},\ }\href {\doibase
  https://urldefense.com/v3/__https://doi.org/10.1016/j.aop.2008.03.004__;!!DZ3fjg!5grOpVkLmw035iyc2zo3Vk1uEzqxInONrgks9mACz0S3pyNin-YA-o3dQpbWkdxaaBv6N0Pdx515iZSrHc4Hv2gkUpZfAQ$}
  {\bibfield  {journal} {\bibinfo  {journal} {Annals of Physics}\ }\textbf
  {\bibinfo {volume} {323}},\ \bibinfo {pages} {2952} (\bibinfo {year}
  {2008})}\BibitemShut {NoStop}%
\bibitem [{\citenamefont {Werner}\ and\ \citenamefont
  {Castin}(2012)}]{Werner2012}%
  \BibitemOpen
  \bibfield  {author} {\bibinfo {author} {\bibfnamefont {F.}~\bibnamefont
  {Werner}}\ and\ \bibinfo {author} {\bibfnamefont {Y.}~\bibnamefont
  {Castin}},\ }\href {\doibase 10.1103/PhysRevA.86.013626} {\bibfield
  {journal} {\bibinfo  {journal} {Phys. Rev. A}\ }\textbf {\bibinfo {volume}
  {86}},\ \bibinfo {pages} {013626} (\bibinfo {year} {2012})}\BibitemShut
  {NoStop}%
\bibitem [{\citenamefont {Gubernatis}\ \emph {et~al.}(2017)\citenamefont
  {Gubernatis}, \citenamefont {Kawashima},\ and\ \citenamefont
  {Werner}}]{Gubernatis2017}%
  \BibitemOpen
  \bibfield  {author} {\bibinfo {author} {\bibfnamefont {J.~E.}\ \bibnamefont
  {Gubernatis}}, \bibinfo {author} {\bibfnamefont {N.}~\bibnamefont
  {Kawashima}}, \ and\ \bibinfo {author} {\bibfnamefont {P.}~\bibnamefont
  {Werner}},\ }\href@noop {} {\emph {\bibinfo {title} {Quantum Monte Carlo
  Methods}}}\ (\bibinfo  {publisher} {Cambridge University Press},\ \bibinfo
  {year} {2017})\BibitemShut {NoStop}%
\bibitem [{\citenamefont {Dean}\ \emph {et~al.}(1993)\citenamefont {Dean},
  \citenamefont {Koonin}, \citenamefont {Lang}, \citenamefont {Ormand},\ and\
  \citenamefont {Radha}}]{Dean1993}%
  \BibitemOpen
  \bibfield  {author} {\bibinfo {author} {\bibfnamefont {D.}~\bibnamefont
  {Dean}}, \bibinfo {author} {\bibfnamefont {S.}~\bibnamefont {Koonin}},
  \bibinfo {author} {\bibfnamefont {G.}~\bibnamefont {Lang}}, \bibinfo {author}
  {\bibfnamefont {W.}~\bibnamefont {Ormand}}, \ and\ \bibinfo {author}
  {\bibfnamefont {P.}~\bibnamefont {Radha}},\ }\href {\doibase
  https://urldefense.com/v3/__https://doi.org/10.1016/0370-2693(93)90995-T__;!!DZ3fjg!5grOpVkLmw035iyc2zo3Vk1uEzqxInONrgks9mACz0S3pyNin-YA-o3dQpbWkdxaaBv6N0Pdx515iZSrHc4Hv2iE9p0HnQ$}
  {\bibfield  {journal} {\bibinfo  {journal} {Physics Letters B}\ }\textbf
  {\bibinfo {volume} {317}},\ \bibinfo {pages} {275} (\bibinfo {year}
  {1993})}\BibitemShut {NoStop}%
\bibitem [{\citenamefont {Ormand}\ \emph {et~al.}(1994)\citenamefont {Ormand},
  \citenamefont {Dean}, \citenamefont {Johnson}, \citenamefont {Lang},\ and\
  \citenamefont {Koonin}}]{Ormand1994}%
  \BibitemOpen
  \bibfield  {author} {\bibinfo {author} {\bibfnamefont {W.~E.}\ \bibnamefont
  {Ormand}}, \bibinfo {author} {\bibfnamefont {D.~J.}\ \bibnamefont {Dean}},
  \bibinfo {author} {\bibfnamefont {C.~W.}\ \bibnamefont {Johnson}}, \bibinfo
  {author} {\bibfnamefont {G.~H.}\ \bibnamefont {Lang}}, \ and\ \bibinfo
  {author} {\bibfnamefont {S.~E.}\ \bibnamefont {Koonin}},\ }\href {\doibase
  10.1103/PhysRevC.49.1422} {\bibfield  {journal} {\bibinfo  {journal} {Phys.
  Rev. C}\ }\textbf {\bibinfo {volume} {49}},\ \bibinfo {pages} {1422}
  (\bibinfo {year} {1994})}\BibitemShut {NoStop}%
\bibitem [{\citenamefont {Gilbreth}\ and\ \citenamefont
  {Alhassid}(2015)}]{Gilbreth2015}%
  \BibitemOpen
  \bibfield  {author} {\bibinfo {author} {\bibfnamefont {C.}~\bibnamefont
  {Gilbreth}}\ and\ \bibinfo {author} {\bibfnamefont {Y.}~\bibnamefont
  {Alhassid}},\ }\href {\doibase
  https://urldefense.com/v3/__https://doi.org/10.1016/j.cpc.2014.09.002__;!!DZ3fjg!5grOpVkLmw035iyc2zo3Vk1uEzqxInONrgks9mACz0S3pyNin-YA-o3dQpbWkdxaaBv6N0Pdx515iZSrHc4Hv2gE1MbMGw$}
  {\bibfield  {journal} {\bibinfo  {journal} {Computer Physics Communications}\
  }\textbf {\bibinfo {volume} {188}},\ \bibinfo {pages} {1} (\bibinfo {year}
  {2015})}\BibitemShut {NoStop}%
\bibitem [{\citenamefont {Gilbreth}\ \emph {et~al.}(2021)\citenamefont
  {Gilbreth}, \citenamefont {Jensen},\ and\ \citenamefont
  {Alhassid}}]{Gilbreth2021}%
  \BibitemOpen
  \bibfield  {author} {\bibinfo {author} {\bibfnamefont {C.~N.}\ \bibnamefont
  {Gilbreth}}, \bibinfo {author} {\bibfnamefont {S.}~\bibnamefont {Jensen}}, \
  and\ \bibinfo {author} {\bibfnamefont {Y.}~\bibnamefont {Alhassid}},\
  }\href@noop {} {\bibfield  {journal} {\bibinfo  {journal} {Computer Physics
  Communications}\ }\textbf {\bibinfo {volume} {264}} (\bibinfo {year}
  {2021})}\BibitemShut {NoStop}%
\bibitem [{\citenamefont {He}\ \emph {et~al.}(2019)\citenamefont {He},
  \citenamefont {Shi},\ and\ \citenamefont {Zhang}}]{He2019}%
  \BibitemOpen
  \bibfield  {author} {\bibinfo {author} {\bibfnamefont {Y.-Y.}\ \bibnamefont
  {He}}, \bibinfo {author} {\bibfnamefont {H.}~\bibnamefont {Shi}}, \ and\
  \bibinfo {author} {\bibfnamefont {S.}~\bibnamefont {Zhang}},\ }\href
  {\doibase 10.1103/PhysRevLett.123.136402} {\bibfield  {journal} {\bibinfo
  {journal} {Phys. Rev. Lett.}\ }\textbf {\bibinfo {volume} {123}},\ \bibinfo
  {pages} {136402} (\bibinfo {year} {2019})}\BibitemShut {NoStop}%
\bibitem [{\citenamefont {Nightingale}(1982)}]{Nightingale1982}%
  \BibitemOpen
  \bibfield  {author} {\bibinfo {author} {\bibfnamefont {P.}~\bibnamefont
  {Nightingale}},\ }\href {\doibase 10.1063/1.330232} {\bibfield  {journal}
  {\bibinfo  {journal} {Journal of Applied Physics}\ }\textbf {\bibinfo
  {volume} {53}},\ \bibinfo {pages} {7927} (\bibinfo {year}
  {1982})}\BibitemShut {NoStop}%
\bibitem [{\citenamefont {dos Santos}\ and\ \citenamefont
  {Sneddon}(1981)}]{Santos1981}%
  \BibitemOpen
  \bibfield  {author} {\bibinfo {author} {\bibfnamefont {R.~R.}\ \bibnamefont
  {dos Santos}}\ and\ \bibinfo {author} {\bibfnamefont {L.}~\bibnamefont
  {Sneddon}},\ }\href {\doibase 10.1103/PhysRevB.23.3541} {\bibfield  {journal}
  {\bibinfo  {journal} {Phys. Rev. B}\ }\textbf {\bibinfo {volume} {23}},\
  \bibinfo {pages} {3541} (\bibinfo {year} {1981})}\BibitemShut {NoStop}%
\bibitem [{sup()}]{supp}%
  \BibitemOpen
  \href@noop {} {}\bibinfo {note} {See the Supplemental Material accompanying
  this article}\BibitemShut {NoStop}%
\bibitem [{\citenamefont {Carlson}\ \emph {et~al.}(2003)\citenamefont
  {Carlson}, \citenamefont {Chang}, \citenamefont {Pandharipande},\ and\
  \citenamefont {Schmidt}}]{Carlson2003}%
  \BibitemOpen
  \bibfield  {author} {\bibinfo {author} {\bibfnamefont {J.}~\bibnamefont
  {Carlson}}, \bibinfo {author} {\bibfnamefont {S.-Y.}\ \bibnamefont {Chang}},
  \bibinfo {author} {\bibfnamefont {V.~R.}\ \bibnamefont {Pandharipande}}, \
  and\ \bibinfo {author} {\bibfnamefont {K.~E.}\ \bibnamefont {Schmidt}},\
  }\href {\doibase 10.1103/PhysRevLett.91.050401} {\bibfield  {journal}
  {\bibinfo  {journal} {Phys. Rev. Lett.}\ }\textbf {\bibinfo {volume} {91}},\
  \bibinfo {pages} {050401} (\bibinfo {year} {2003})}\BibitemShut {NoStop}%
\bibitem [{\citenamefont {Gezerlis}\ and\ \citenamefont
  {Carlson}(2008)}]{Gezerlis2008}%
  \BibitemOpen
  \bibfield  {author} {\bibinfo {author} {\bibfnamefont {A.}~\bibnamefont
  {Gezerlis}}\ and\ \bibinfo {author} {\bibfnamefont {J.}~\bibnamefont
  {Carlson}},\ }\href {\doibase 10.1103/PhysRevC.77.032801} {\bibfield
  {journal} {\bibinfo  {journal} {Phys. Rev. C}\ }\textbf {\bibinfo {volume}
  {77}},\ \bibinfo {pages} {032801} (\bibinfo {year} {2008})}\BibitemShut
  {NoStop}%
\bibitem [{\citenamefont {Jensen}\ \emph
  {et~al.}(2020{\natexlab{b}})\citenamefont {Jensen}, \citenamefont
  {Gilbreth},\ and\ \citenamefont {Alhassid}}]{Jensen2020b}%
  \BibitemOpen
  \bibfield  {author} {\bibinfo {author} {\bibfnamefont {S.}~\bibnamefont
  {Jensen}}, \bibinfo {author} {\bibfnamefont {C.~N.}\ \bibnamefont
  {Gilbreth}}, \ and\ \bibinfo {author} {\bibfnamefont {Y.}~\bibnamefont
  {Alhassid}},\ }\href {\doibase 10.1103/PhysRevLett.125.043402} {\bibfield
  {journal} {\bibinfo  {journal} {Phys. Rev. Lett.}\ }\textbf {\bibinfo
  {volume} {125}},\ \bibinfo {pages} {043402} (\bibinfo {year}
  {2020}{\natexlab{b}})}\BibitemShut {NoStop}%
\end{thebibliography}

\begin{thebibliography}{99}
\bibitem{Nightingale1982S}   P. Nightingale, Journal of Applied Physics {\bf 53}, 7927 (1982).
\bibitem{Santos1981S} R. R. dos Santos and L. Sneddon, Phys. Rev. B {\bf 23}, 3541 (1981).
\bibitem{Moreo1991S} A. Moreo and D. J. Scalapino, Phys. Rev. Lett. {\bf 66}, 946 (1991).
\bibitem{Kosterlitz1973S} J. M. Kosterlitz and D. J. Thouless, Journal of Physics C: Solid State Physics {\bf 6}, 1181 (1973).
\bibitem{Paiva2004S} T. Paiva, R. R. dos Santos, R. T. Scalettar, and P. J. H. Denteneer, Phys. Rev. B {\bf 69}, 184501 (2004).
\bibitem{Bertaina2011S} G. Bertaina and S. Giorgini, Phys. Rev. Lett. {\bf 106}, 110403 (2011).
\bibitem{Galea2016S} A. Galea, H. Dawkins, S. Gandolfi, and A. Gezerlis, Phys. Rev. A {\bf 93}, 023602 (2016).
\bibitem{Vitali2017S} E. Vitali, H. Shi, M. Qin, and S. Zhang, Phys. Rev. A {\bf 96}, 061601 (2017).
\bibitem{Tan2008S} S. Tan, Annals of Physics {\bf 323}, 2952 (2008).
\bibitem{Werner2012S} F. Werner and Y. Castin, Phys. Rev. A {\bf 86}, 013626  (2012).
\end{thebibliography}

\end{document}